\documentclass[draft]{agujournal2019}
\usepackage{url} 
\usepackage{lineno}
\usepackage[inline]{trackchanges} 
\usepackage{soul}
\usepackage{longtable}
\usepackage{multirow}
\usepackage{amsmath}
\usepackage{arydshln}
\nolinenumbers
\usepackage{graphicx}
\draftfalse

\journalname{JGR: Space Physics}

\begin{document}

%
%


\title{Long-Term Density Trend in the Mesosphere and Lower Thermosphere from Occultations of the Crab Nebula with X-Ray Astronomy Satellites}

%
%




\authors{Satoru Katsuda\affil{1}, Teruaki Enoto\affil{2,3}, Andrea N. Lommen\affil{4}, Koji Mori\affil{5,6}, Yuko Motizuki\affil{7}, Motoki Nakajima\affil{8}, Nathaniel C. Ruhl\affil{4}, Kosuke Sato\affil{1}, Gunter Stober\affil{9}, Makoto S. Tashiro\affil{6,1}, Yukikatsu Terada\affil{1,6}, Kent S. Wood\affil{10}}

\affiliation{1}{Graduate School of Science and Engineering, Saitama University, 255 Shimo-Ohkubo, Sakura, Saitama 338-8570, Japan}
\affiliation{2}{Department of Physics, Graduate School of Science, Kyoto University, Kitashirakawa Oiwake-cho, Sakyo-ku, Kyoto 606-8502, Japan}
\affiliation{3}{RIKEN Cluster for Pioneering Research, Hirosawa 2-1, Wako 351-0198, Japan}
\affiliation{4}{Haverford College, 370 Lancaster Ave., Haverford, PA 19041, USA}
\affiliation{5}{Department of Applied Physics and Electronic Engineering, University of Miyazaki, 1-1, Gakuen Kibanadai-nishi, Miyazaki 889-2192, Japan}
\affiliation{6}{Institute of Space and Astronautical Science (ISAS), Japan Aerospace Exploration Agency (JAXA), 3-1-1 Yoshinodai, Chuo-ku, Sagamihara, Kanagawa, 252-5210, Japan}
\affiliation{7}{RIKEN Nishina Center, Hirosawa 2-1, Wako 351-0198, Japan}
\affiliation{8}{School of Dentistry at Matsudo, Nihon University, 2-870-1 Sakaecho-nishi, Matsudo, Chiba 101-8308, Japan}
\affiliation{9}{Institute of Applied Physics \& Oeschger Center for Climate Change Research, Microwave Physics, University of Bern, Bern, Switzerland}
\affiliation{10}{P.O.\ Box 127, Nathrop, Colorado 81236, USA}




\correspondingauthor{Satoru Katsuda}{katsuda@mail.saitama-u.ac.jp}




\begin{keypoints}
\item Time series of combined O and N densities are measured in the MLT, based on atmospheric occultations of the Crab Nebula using X-ray astronomy satellites.  
\item The density is decreasing everywhere, with a local minimum of $-$12\%/decade near 105\,km. 
\item The local minimum in density trends may be due to strong cooling by water vapor and ozone, as was first predicted by \citeA{2006JASTP..68.1879A}.
\end{keypoints}

%
%

%
%


\begin{abstract}

We present long-term density trends of the Earth's upper atmosphere at altitudes between 71 and 116\,km, based on atmospheric occultations of the Crab Nebula observed with X-ray astronomy satellites, ASCA, RXTE, Suzaku, NuSTAR, and Hitomi.  The combination of the five satellites provides a time period of 28\,yr from 1994 to 2022.  To suppress seasonal and latitudinal variations, we concentrate on the data taken in autumn ($49<$ doy $<111$) and spring ($235<$ doy $<297$) in the northern hemisphere with latitudes of 0$^\circ$--40$^\circ$.  With this constraint, local times are automatically limited either around noon or midnight.  We obtain four sets (two seasons $\times$ two local times) of density trends at each altitude layer.  We take into account variations due to a linear trend and the 11-yr solar cycle using linear regression techniques.  Because we do not see significant differences among the four trends, we combine them to provide a single vertical profile of trend slopes.  We find a negative density trend of roughly $-$5\%/decade at every altitude.   This is in reasonable agreement with inferences from settling rate of the upper atmosphere.  In the 100--110\,km altitude, we found an exceptionally high density decline of about $-$12\%/decade.  This peak may be the first observational evidence for strong cooling due to water vapor and ozone near 110\,km, which was first identified in a numerical simulation by \citeA{2006JASTP..68.1879A}.  Further observations and numerical simulations with suitable input parameters are needed to establish this feature. 

\end{abstract}

\section*{Plain Language Summary}
Numerical simulations have shown that, while an increase of greenhouse gases such as CO$_2$ in the atmosphere causes heating of the troposphere (near surface), it causes cooling of the middle and upper atmosphere, which is the so-called ``greenhouse cooling".  The greenhouse cooling should result in atmospheric contraction and consequently a temporal density decrease at a fixed height.  However, observational evidence for the density decrease has been scarce in the mesosphere and lower thermosphere (MLT: 80--110\,km), owing to difficulty in measuring the density in this region.  Here, we present the first direct measurements of long-term variations for combined N and O atom number density in the MLT, based on atmospheric occultations of the Crab Nebula observed with X-ray astronomy satellites.  The combination of five X-ray astronomy satellites, ASCA, RXTE, Suzaku, NuSTAR, and Hitomi, allows us to explore density trends for a long period from 1994 to 2022.  We take into account variations due to a temporal linear trend and the 11-yr solar cycle, using linear regression techniques.  As a result, we find a negative density trend of roughly $-5$\%/decade at every altitude, with a local minimum of $-12$\%/decade near 105 km.  This is in reasonable agreement with the state-of-the-art numerical simulations.

%
%

\section{Introduction}

Understanding the changes of the middle and upper atmosphere is important to our life, as was illustrated by the discovery of the stratospheric ozone depletion.  The importance is increasing, as we utilize the space environment for many purposes such as telecommunications, navigation, space tourism, and scientific projects to explore the universe.  There are several known drivers of middle and upper atmosphere climate change, including the increase in greenhouse gas (mainly, water vapour and carbon dioxide) concentrations \cite<e.g.,>{2006JASTP..68.1879A}, changes in ozone concentration \cite<e.g.,>{2008JASTP..70.1473B}, long-term solar and geomagnetic activity variations \cite<e.g.,>{2011JGRA..116.0H12B,2019JGRA..124.6305P}, and shifts in Earth's main magnetic field \cite<e.g.,>{2016JGRD..121.7781C}.  Of these, the dominant cause of long-term trends in the upper atmosphere in recent decades is thought to be the increase of greenhouse gases \cite<e.g.,>{2006JASTP..68.1879A,2013JGRA..118.3846Q,2020JGRA..12528623C}.  

The roles of greenhouse gases in the atmosphere are (1) heating the atmosphere in the troposphere (near surface) and (2) cooling the middle atmosphere above $\sim$20\,km altitude, which was first shown by computer simulations by \citeA{1967JAtS...24..241M}.  \citeA{1989GeoRL..16.1441R} performed pioneering numerical simulations focusing on the middle and upper atmosphere above 60\,km, followed by many researchers in both numerical simulations and observations \cite<see, e.g.,>[for recent reviews]{2017JASTP.163....2L,2020Ge&Ae..60..397D}.  Greenhouse gases, particularly CO$_2$, cool the middle atmosphere, because the density in the middle atmosphere is so low that CO$_2$ molecules (and all the other greenhouse gases) are optically thin to outgoing infrared radiation.  In this condition, thermal energy transferred by collisions with ambient gas to the excited states of CO$_2$ is lost to space via its infrared radiation \cite<e.g.,>[for a deeper understanding of CO$_2$-induced cooling]{2016ESD.....7..697G}.  It is the enhancement of greenhouse cooling, via increases in CO$_2$ and other greenhouse gases, that changes the thermal balance and induces the long-term trends. The combination of lower atmosphere heating and middle atmosphere cooling is demonstrated on Venus, whose atmospheric composition is dominated by CO$_2$.  The ``greenhouse cooling" in the middle atmosphere results in contraction of the middle atmosphere, leading to a downward displacement of constant pressure surfaces \cite<e.g.,>{2006JASTP..68.1879A}.  Combined with the exponential density decrease with increasing altitude, a density at a fixed height decreases with time.  The fact that stronger long-term trends have been predicted and confirmed in the middle atmosphere rather than the troposphere makes studies of the mesosphere and lower thermosphere (MLT) important in predicting climate changes.

The density trends in the thermosphere ($\sim$300\,km altitude and above) have been extensively studied from satellites' orbit drag.  Various researchers came to the conclusion that the density trend at $\sim$400\,km is negative at about a few \%/decade with a possible solar-cycle dependence, ranging from $-5$\%/decade for solar minimum conditions to $-2$\%/decade for maximum conditions \cite<e.g.,>[and references therein]{2008GeoRL..35.5101E}.  The latest results as well as reviews of previous work can be found in \citeA{2015JGRA..120.2940E}.  The observational trend is in reasonable agreement with the state-of-the-art numerical simulations \cite<e.g.,>{2015JGRA..120.2183S,2020JGRA..12528623C}.  

The trend study in the MLT (80--110\,km altitude) is also of great importance, especially because volume mixing ratio (VMR) of CO$_2$ rapidly falls off in this region due to molecular diffusive separation (as CO$_2$ is heavier than the mean molecular weight of the air) and photolysis.  According to Whole Atmosphere Community Climate Model (WACCM) simulations \cite<e.g.,>{2017JGRD..122.8345L}, the CO$_2$ VMR decreases from 380\,ppmv below 80\,km to 120\,ppmv at 110\,km altitude.  Observationally, the CO$_2$ VMR is relatively difficult to be quantified in this region \cite[for a review of observations]{2019JASTP.189...80L}.  The most recent measurements are consistent with the model at the bottom of the MLT, but depart from the model with increasing altitude in the sense that the model underestimates the data \cite<e.g.,>{2016JGRD..121.3634G}, posing challenges to our current understanding of dynamics, energetics, and photolysis in the MLT region.  The long-term trend of the air density in the MLT region can be a new indicator of the trend of the CO$_2$ VMR trend.  

Despite its importance, measurements of density trends in the MLT region are still scarce due to difficulty in in-situ observations.  Snap-shot measurements were performed from solar/stellar occultations \cite<e.g.,>{1968JGR....73.5798N,2007JGRA..112.6323D} as well as in-situ measurements based on the falling sphere technique \cite<e.g.,>[and references therein]{2013aisi.conf....1S}.  Although these data provided precious opportunities to verify density models, they are not sufficient to explore long-term trends.  There is only one indirect measurement of the long-term density trends in the MLT.  This is based on meteor radar-echo observations which measure meteor peak flux altitudes and convert them to neutral air densities.  The neutral density trend inferred by this method was $-5.8\pm1.1$\%/decade at 91\,km altitude \cite{2014GeoRL..41.6919S}.  Because this is obtained at a single altitude, our knowledge of the long-term trend in the MLT region is obviously poor.  

Here, we present the first direct measurements of long-term neutral density variations in the MLT (altitudes between 71 and 116\,km), utilizing atmospheric occultations of the Crab Nebula observed with X-ray astronomy satellites.  The X-ray occultation method allows us to measure atomic number densities integrated along the line of sight, i.e., the column densities.  Because X-rays are directly absorbed by inner K-shell and L-shell electrons, X-rays see only atoms (within molecules).  Thus, the X-ray occultation method can not distinguish between atoms and molecules, but we can obtain their total number density without complexity involved with chemical, ionization, and excitation processes.  This is different from other methods of density measurements such as orbital drag which measures mass density, meteor radar which also effectively measures mass density when converting height to density, optical and infrared remote-sensing techniques like TIMED/SABER \cite{1999SPIE.3756..277R} which typically infer the atmospheric pressure and temperature, which are easily convertible via the ideal gas law to the number density.  More specifically, the X-ray occultation method allows us to derive column densities of combined N and O atoms, i.e., $N_{\rm A,N+O} = 2$$\times$$N_{\rm M,N_2} + 2$$\times$$N_{\rm M,O_2} + N_{\rm A,O}$, where subscripts A and M represent the atomic number density and molecular number density, respectively (other minor species such as NO, N, and OH do not contribute significantly to X-ray extinction).  We can not distinguish between N and O, because the energy dependence of photon cross sections of N and O are similar with each other in the energy range of 1--100\,keV.

A pioneering work on X-ray occultation sounding of the upper atmosphere was performed by \citeA{2007JGRA..112.6323D} who analyzed Earth's atmospheric occultations of the Crab Nebula obtained with the RXTE satellite.  Historically, this paper was produced as a by-product of X-ray satellite navigation experiments; if the navigational state vector for the satellite is known with precision, then the horizon crossings can be used to extract diagnostic information about the state of the atmosphere.  The history of the dual usage of X-ray horizon crossing transitions is summarized in \citeA{Wood2020}.  Recently, \citeA{2021JGRA..12628886K}, hereafter Paper~I, and \citeA{2022AdSpR..69.3426Y,2022AMT....15.3141Y,2023ApJS..264....5X} applied the same technique to derive atmospheric vertical density profiles, using Japanese and Chinese X-ray astronomy satellites, respectively.  In this paper, we analyze atmospheric occultation data of the Crab Nebula acquired with X-ray astronomy satellites, ASCA, RXTE, Suzaku, NuSTAR, and Hitomi to reveal long-term trend of the neutral density in the MLT region.  The data were obtained between 1994 and 2022, providing a time span of 28\,yr.  This is longer than two solar cycles, enabling us to remove/suppress the possible influence of the 11-year solar cycle.  This paper is organized as follows.  In Section 2, we describe observations analyzed in this paper.  In Section 3, we describe data analysis and results.  In Section 4, we interpret the analysis results.  Finally, we give conclusions of this paper in Section 5.

\section{Observations and Data Reduction}

We analyze X-ray data during Earth occultations of the Crab Nebula, which is a remnant of a supernova explosion that occurred in 1054~AD, to obtain vertical density profiles of the MLT region.  The essential requirements for the celestial source used for this technique are (1) that the source be either constant or not varying significantly during the time required for the atmospheric occultation (a minute or so) and (2) that it should be a point source or sufficiently point-like that the angular extent does not matter.  It is known that the Crab Nebula is not a perfect standard candle \cite{2011ApJ...727L..40W}.  However, its degree of variability is not so rapid as to change the shapes of individual occultations, or affect the analysis.  It is also well established that the Crab Nebula also has an angular extent of about an arcminute in X-rays around 1\,keV and then decreases in angular extent with increasing X-ray energy.  Given that the distance between the satellite and the tangent point is roughly 2000\,km, this angular extent corresponds to a tangent altitude extent of $\sim$0.6\,km the tangent point is where the line of sight is closest to the Earth and the tangent altitude is the height from the Earth at the tangent point.  This is comparable with the error due to the satellite position uncertainty from two-line element (TLE: Data are accessible by sending special requests at {\tt http://celestrak.com/NORAD/archives/request.php}) $+$ Simplified General Perturbations Satellite Orbit Model 4 (SGP4), as we will quantify later in this section.  One could worry from these rough estimates that the angular size of the Crab Nebula, which is slightly asymmetric, should be in the error analysis, because it might be of the same order as the satellite position error.  However, while there is potential for the total angular extent to be a problem, the flux is concentrated toward the center of the Nebula and the pulsar contributes as a point source near the centroid of the Nebular emission \cite<e.g., Figs.~11--15 in>{2015ApJ...801...66M}.  Thus, it is reasonable to consider the Crab Nebula as effectively a point source.  

\begin{figure}
\includegraphics[width=35pc,angle=0]{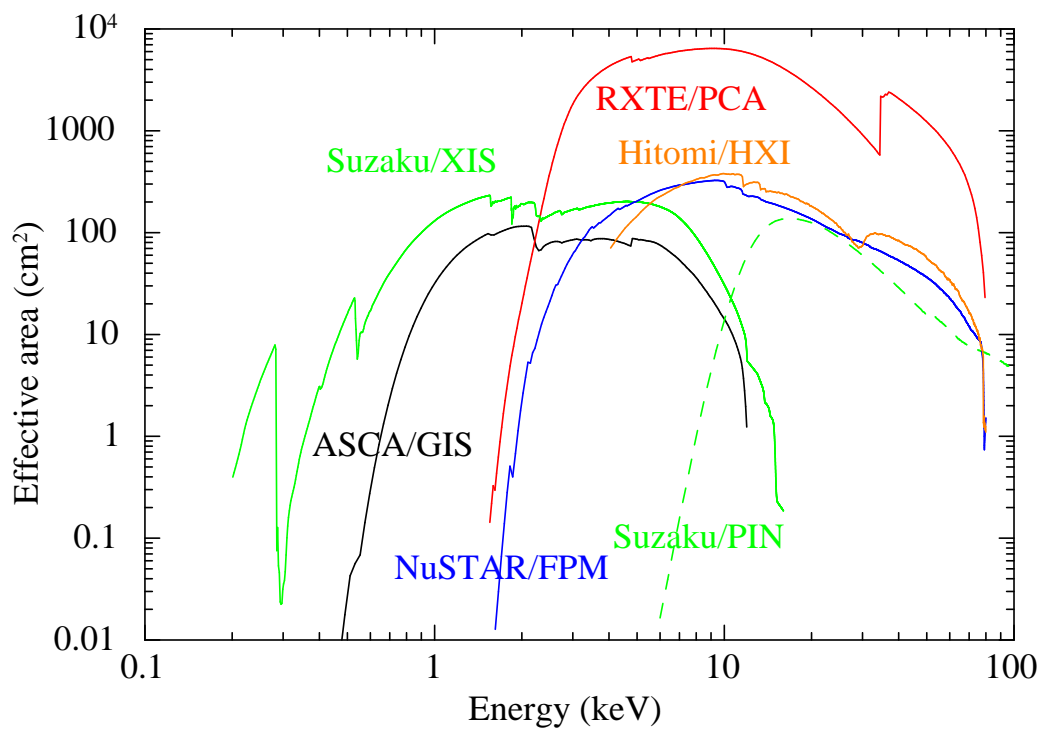}
\caption{Comparison of the effective areas for the six instruments analyzed in this paper.}
\label{fig:all_eff}
\end{figure}

Basic information about the X-ray occultation technique can be found in Paper~I.  Because most (if not all) X-ray astronomy satellites so far have used the Crab Nebula to perform their calibrations, there are numerous data sets for this standard candle on orbits.  Here, we make use of them to reveal long-term trends of the air density.  Specifically, we use data acquired with the X-ray astronomy satellites/instruments, ASCA/GIS \cite{1994PASJ...46L..37T,1996PASJ...48..157O}, RXTE/PCA \cite{1996SPIE.2808...59J}, Suzaku/XIS+PIN \cite{2007PASJ...59S...1M,2007PASJ...59S..23K,2007PASJ...59S..35T}, NuSTAR/FPM \cite{2013ApJ...770..103H}, and Hitomi/HXI \cite{2018JATIS...4b1402T,2018JATIS...4b1410N}.  Capabilities of individual instruments are summarized in Table~\ref{tab:instruments}.  The effective areas are plotted in Fig.~\ref{fig:all_eff}.

To remove possible seasonal biases, we concentrate on spring ($\pm$30 days around the spring equinox) and autumn ($\pm$30 days around the autumnal equinox), and analyze the two seasons separately.  In fact, ASCA, Suzaku, and Hitomi observed the Crab Nebula only in these two seasons, whereas RXTE and NuSTAR observed it during the whole year.  Given that the five X-ray satellites used in this paper are in the low Earth orbits with low inclination angles, and that the declination angle of the Crab Nebula is $+$22$^\circ$ above the celestial equator, the majority of the data are obtained in the northern hemisphere.  To eliminate/suppress possible latitudinal dependences of atmospheric density, we analyze data obtained only in the northern hemisphere.  Consequently, all the data are acquired in a latitude range between 0$^\circ$ and 40$^\circ$ in the northern hemisphere.  Because this latitude range is large, it would be ideal to divide the data even more finely by latitude.  However, the limited number of data makes such an analysis difficult.  Alternatively, we could reduce the latitude-dependent density variations by analyzing deviations from an appropriate model such as NRLMSIS 2.0.  Although such a fine tuning will not be considered in the present work, it will be worthwhile in the future analysis.

\begin{figure}
\includegraphics[width=35pc,angle=0]{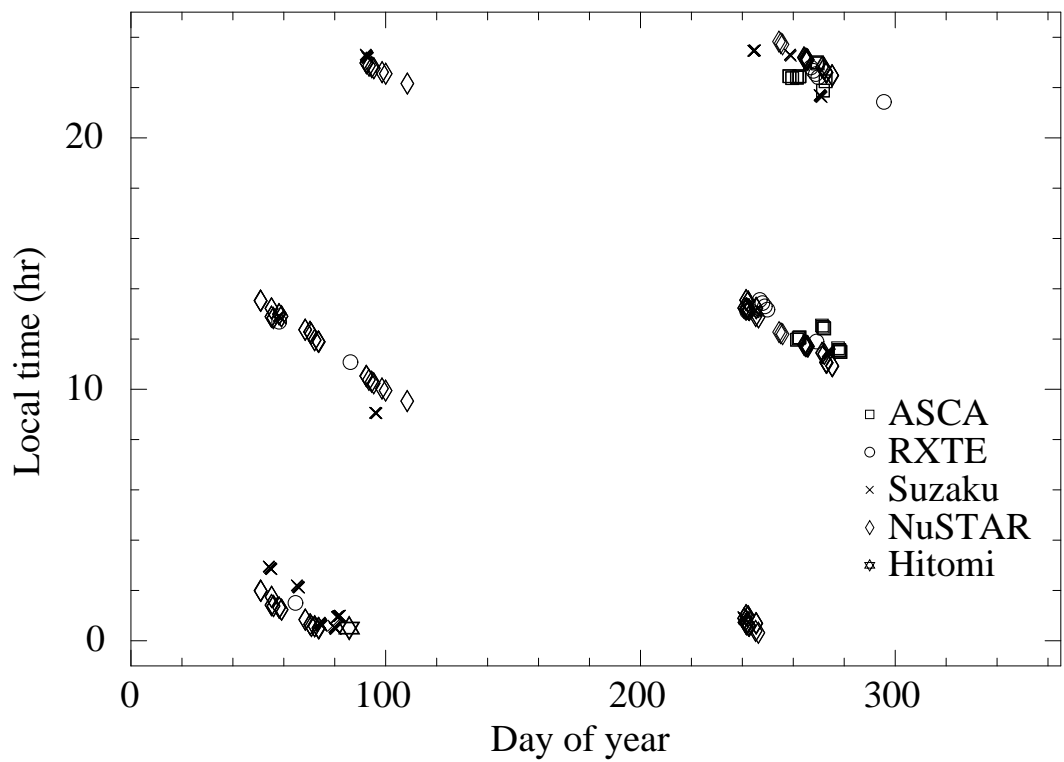}
\caption{Local time at an altitude of 100\,km for each occultation against the day of year.}  
\label{fig:obs_loc}
\end{figure}

In Table~\ref{tab:data}, we summarize basic information about all the data analyzed in this paper.  The individual setting (or rising) occultations in the same observation sequences are combined together to improve the photon statistics.  This merging process is reasonable, because these data are taken at the similar telescope tangent points (latitude and longitude) and local times with each other.  The merged occultations are labeled with the same group number as shown in the last column of the table.  We note that the photon statistics in the RXTE data are so rich that they need not be combined.  From this table and Fig.~\ref{fig:obs_loc}, which illustrates local time vs.\ day of year for all the occultation scans analyzed, we can see that local times are approximately either midnight or noon for all of our data sets.  Therefore, by analyzing the noon and midnight data separately, density variations caused by tidal waves with periods of 24\,hr plus harmonics are removed naturally.  

In Paper~I, we did not use the satellite positions provided by the Suzaku and Hitomi housekeeping files, but rather recalculated their positions by using the TLEs to improve accuracies of their positions and consequently tangent altitudes.  In the mean time, we realized that the original positions provided by JAXA, which we received after publishing Paper~I, are actually more accurate than the TLE data and are used in this paper.  Suzaku's position accuracy is not available, but is a few hundred meters, given that it is determined by radio ranging.  This is an order of magnitude better than that for the TLE data, which will be evaluated later in this section.  Because Hitomi carries a GPS (global pointing system) receiver, its position accuracy is less than 1\,m.   The accuracy in the position of the RXTE satellite is less than $\sim$450\,m with 99\% confidence \cite{2006ApJS..163..401J}.  Such accurate positions are recorded every 60 sec.  Unfortunately, this time resolution is insufficient for our analysis.  Therefore, we interpolate the satellite positions by using a cubic spline interpolation method.  The accuracy of the interpolated position becomes worse than the original precision, but is less than a few 100\,m, which is sufficient for the following analysis.  

As for ASCA and NuSTAR, no accurate satellite positions (with reasonable time intervals for ASCA) are available.  Therefore, we calculate their positions based on TLE sets with SGP4, using the {\tt skyfield} software written in python (available at {\tt https://rhodesmill.org/skyfield/}.  The uncertainty on the TLE-based satellite positions are estimated for the RXTE satellite by comparing the TLE-based positions and high precision orbit data provided by the RXTE team.  For this comparison, we do not perform the cubic interpolation but use original data with a cadence of 60\,sec to keep the original high precision.  To increase the number of samples, we collect all occultation data of the Crab Nebula without constraints of the season and the latitude.  As a result, we obtained a total of 140 occultations, most of which are not listed in Table~\ref{tab:data}.  For each occultation, we calculate position differences between TLE and high-precision orbits for about 10 data points ($\sim$10 minutes) including the atmospheric occultation, and calculate their averages.  The difference is calculated in altitude, latitude, and longitude directions, separately.  

\begin{figure}
\includegraphics[width=35pc,angle=0]{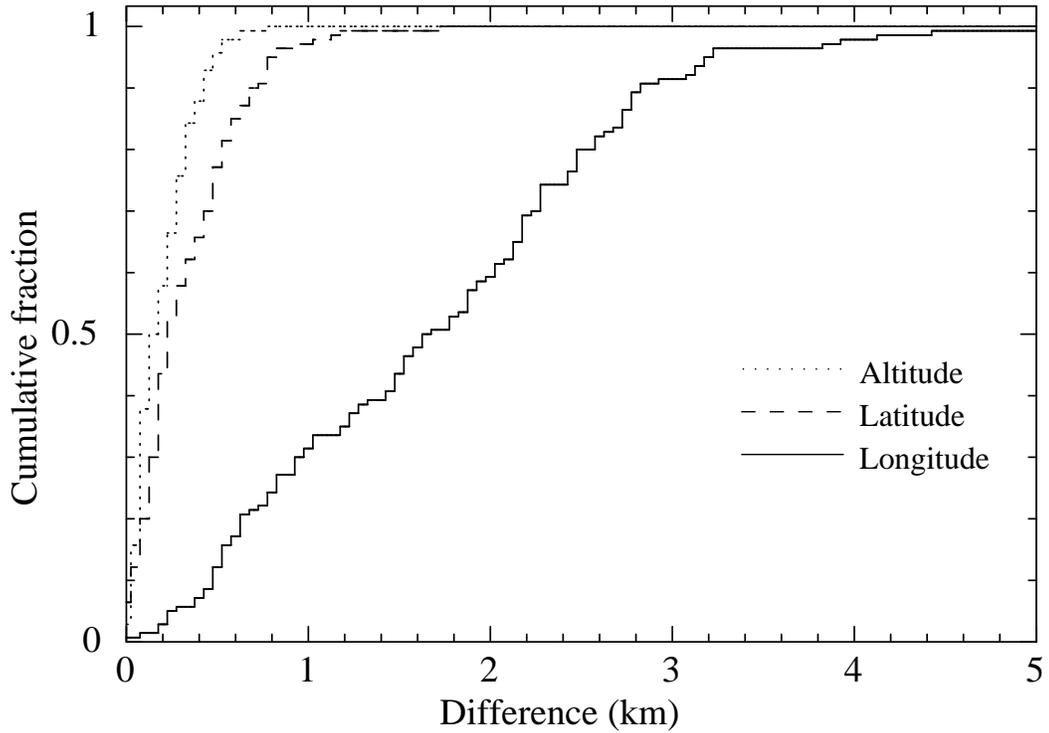}
\caption{Cumulative fraction of the TLE errors in three directions.  The TLE errors are defined as the difference between the RXTE position given in the orbit file (less than $\sim$450-m at 99\% confidence level) and that estimated from the TLE.  }
\label{fig:TLEerror}
\end{figure}

Figure~\ref{fig:TLEerror} shows a cumulative fraction of the position differences in three directions, from which we derive the position uncertainties on the TLE as shown in Table~\ref{tab:TLEerror}.  The fact that errors in the longitudinal direction are much larger than those in the other directions is reasonable, because TLEs tend to have larger errors in the in-track direction \cite{2009AdSpR..43.1065W}, which is close to the longitudinal direction for the RXTE with inclination angle of only 23$^\circ$.  In fact, the orbit errors associated with TLE+SGP4 measured for the CHAMP satellite \cite{2018Ap&SS.363...31X} are close to our measurement for the RXTE satellite.  Thus, the errors in Table~\ref{tab:TLEerror} seem typical for TLE-based positions for low-Earth-orbit satellites.  In the next section, we will evaluate the impact of the spacecraft's position error to the measurement error on the air density.

\section{Analysis and Results}

\begin{figure}
\includegraphics[width=35pc,angle=0]{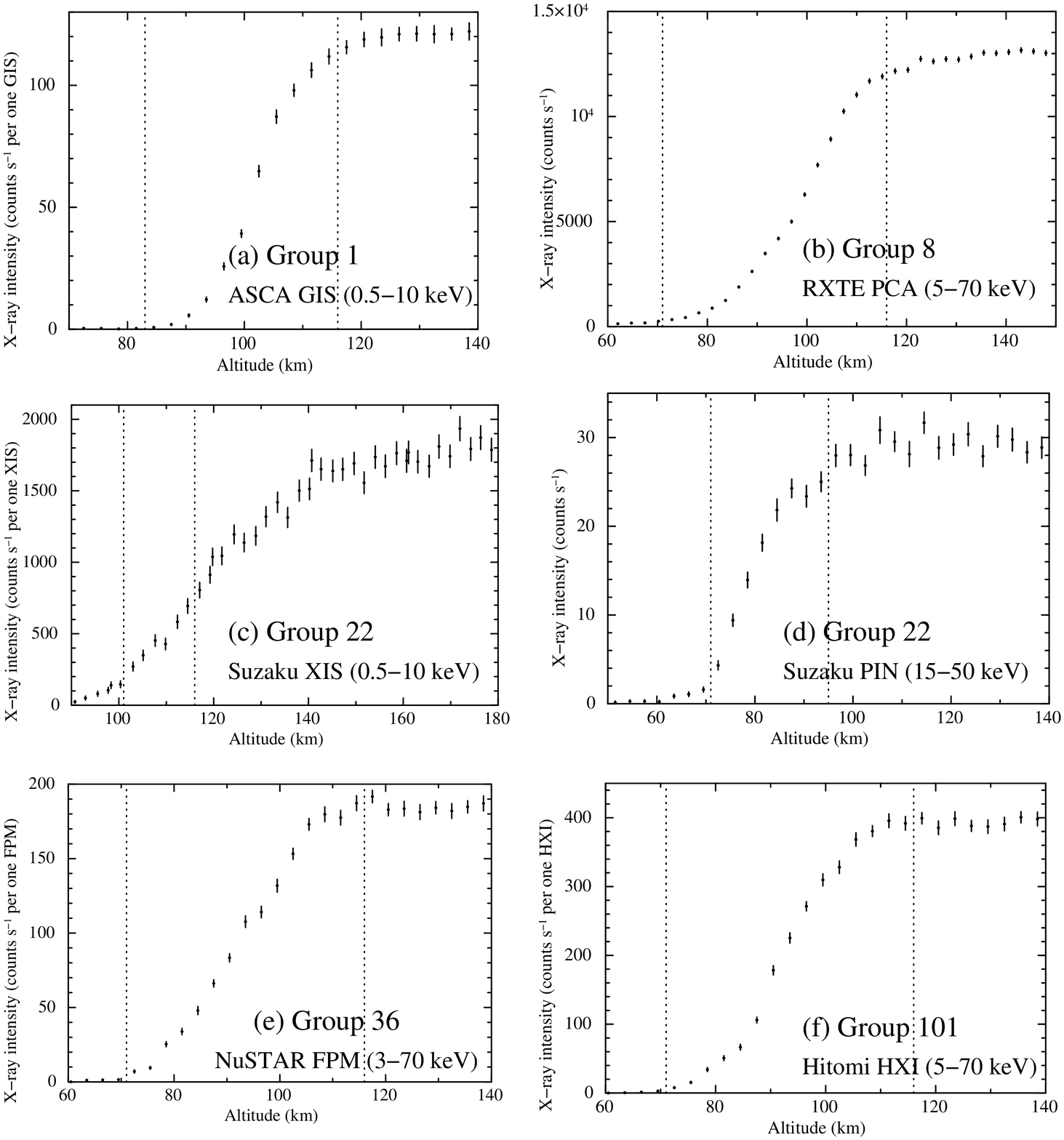}
\caption{(a) Example occultation light curves for ASCA/GIS.  Ten individual occultations in group 1 (five occultations for GIS2 and GIS3) are merged to improve the photon statistics.  The deadtime correction is not performed for this light curve.   The altitude range used for our spectral analysis is within dotted vertical lines in each panel.  (b)--(f) Same as (a) but for RXTE/PCA, Suzaku/XIS, Suzaku/PIN, NuSTAR/FPM, and Hitomi/HXI.  Note that rounded shapes of the light curves are due to gradual changes of the neutral column density as a function of altitude.  Also, the altitudes where most of the photons are absorbed vary significantly.  This is because energy coverages are different among instruments; photons with higher energies can penetrate more deeply into the atmosphere.}
\label{fig:lc}
\end{figure}

We plot in Fig.~\ref{fig:lc} example occultation light curves taken with each of the six instruments used in this paper.  The gradual increase/decrease of the X-ray intensity along with the tangent altitude clearly shows the effect of atmospheric absorption.  Also, the characteristic rising/setting altitudes significantly vary from instrument to instrument.  This is due to the different energy coverage of each instrument, as shown in Fig.~\ref{fig:all_eff} showing the effective area comparison for the six instruments.  The higher the energy coverage of the instrument, the deeper the characteristic altitude.  The altitude range to be analyzed for our spectral analysis is within dotted vertical lines in each panel, which is slightly different from instrument to instrument.

In principle, we can derive the atmospheric density profile by analyzing the light curves by using the Beer's law method \cite{1993JGR....9817607A}.  This method is particularly useful for cases in which emission/absorption lines are analyzed, because one can know the accurate photoelectric cross section at the photon energy of interest.  On the other hand, the X-ray spectrum from the Crab Nebula is a power-law continuum.  In this case, it is not very easy to specify a photon energy for a light curve, leading to a substantial uncertantiy on the cross section.  Therefore, we analyze the X-ray spectrum to derive vertical atmospheric density profiles.  In this case, it is technically easy to take account of the energy-dependent absorption cross section, by utilizing the XSPEC package \cite{1996ASPC..101...17A} which is a standard spectral analysis software in the X-ray astronomy.

We measure vertical density profiles for each of data group defined in Table~\ref{tab:data}.  The procedure is almost the same as we did in Paper~I, but some modifications and improvements are applied as described in the next few paragraphs.  Briefly, we extract X-ray spectra from an altitude layer between 71\,km and 116\,km, with a resolution of 6\,km for each occultation data group.  Every 6-km layer is overlapped with adjacent layers by half to derive smooth density profiles.  An exception is Suzaku/XIS for which the data were obtained in a special model, i.e., an exposure time of 0.1\,sec with a cadence of 8\,sec.  As Suzaku moves in orbit, the tangent altitude increases/decreases at a speed of $\sim$2\,km/sec near a tangent altitude of 100\,km.  Thus, the altitude bin of the XIS data corresponds to $\sim$0.2\,km and the data are taken every $\sim$16\,km altitude.  This is too sparse for binning by 6\,km.  On the other hand, thanks to the high throughput of the combined three co-aligned XISs, the exposure time of 0.1\,sec (or 0.2\,km altitude resolution) is sufficient for us to perform spectral analysis.  Thus, we analyze the XIS spectra without binning.

When observing very bright sources like the Crab Nebula, ASCA/GIS, Suzaku/PIN, and NuSTAR/FPM have substantial deadtimes, i.e., the periods when the system is unable to record another event after the recording of each of new events.  For example, the ASCA/GIS has a typical deadtime of about 8 msec per event \cite{1996PASJ...48..171M}.  Thus, in our spectral analysis, we perform deadtime corrections for ASCA/GIS, Suzaku/PIN, and NuSTAR/FPM, for which deadtime fractions for the Crab Nebula exceed 10\% at the top of the atmosphere.  The correction is simply to recalculate the exposure time, by multipying the exposure time by $1 - f_{\rm DT}$, where $f_{\rm DT}$ is the deadtime fraction.  To calculate deadtime fractions, we utilize {\tt hxddtcor} for Suzaku/PIN.  As for the ASCA/GIS and NuSTAR/FPM, we calculated deadtime fractions from the detected count rates by using Fig.~18 (specifically, the theoretical prediction for Fast Lorentzian Fit SPREAD-ON, or FLF SP-ON) in \citeA{1996PASJ...48..171M} for ASCA/GIS or Fig.~2.7 in \citeA{2012PhDT.........5B} for NuSTAR/FPM, respectively. 

Before measuring the atmospheric density, we define unattenuated spectra from the Crab Nebula, i.e., the source spectrum at the top of the Earth's atmosphere, for each occultation data group.  To this end, we fit X-ray spectra taken during tangential altitudes well above the atmospheric attenuation with a conventional emission model for the Crab Nebula.  The model consists of {\tt TBabs} $\times$ {\tt power-law}, where {\tt TBabs} is the interstellar absorption model \cite{2000ApJ...542..914W} and {\tt power-law} represents the synchrotron radiation from the Crab Nebula.  Then, we evaluate lower-altitude X-ray spectra with a spectral model taking account of the effect of the atmospheric absorption.  In this procedure, our model is expressed as {\tt vphabs} $\times$ {\tt cabs} $\times$ {\tt TBabs} $\times$ {\tt powerlaw}, where {\tt vphabs} and {\tt cabs} represent the photo-electric absorptions and the Compton scattering due to the Earth's atmosphere.  In Paper~I, we did not consider the {\tt cabs} model.  However, the Compton scattering could play an important role in the hard X-ray regime; a total photon cross section in carbon is dominated by the Compton scattering process in the photon energy range of 0.05--10\,MeV \cite{1980JPCRD...9.1023H}.  Thus, in this paper we add it to derive more accurate density profiles especially in the atmosphere below 100\,km altitude.

\begin{figure}
\includegraphics[width=35pc,angle=0]{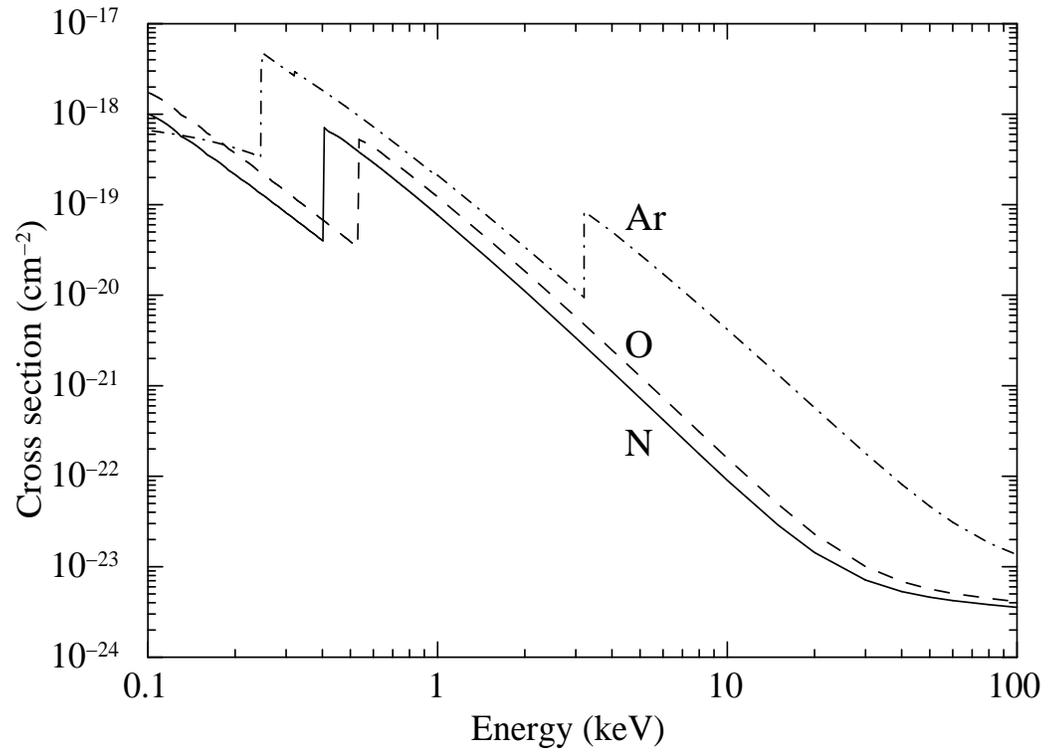}
\caption{ X-ray cross sections for N, O, and Ar atoms.  Solid, dashed, and dash-dotted lines are for N, O, and Ar, respectively.}
\label{fig:Xsect}
\end{figure}

We assume that the atmosphere consists purely of N, O, and Ar, as we did in Paper~I.  The photon cross sections per atom are given in Fig.~\ref{fig:Xsect}, where we took the data below and above 1\,keV from \cite{1996ApJ...465..487V} and {\tt XCOM} (available at {\tt https://physics.nist.gov/PhysRefData/Xcom/html/xcom1.html}), respectively.  Given that the density of Ar is less than $\sim$0.5\% of N, its contribution to the X-ray absorption is so small that we can not measure its column density.  Also, the similarity in cross sections of N and O does not allow us to measure N and O abundances separately.  Therefore, we fix the relative abundances of O/N and Ar/N to model values from NRLMSIS 2.0 \cite{2021E&SS....801321E}.  We calculate the relative abundances of N, O, and Ar atoms from the MSIS number densities of N$_2$, O$_2$, O, and Ar, by utilizing a python package {\tt pymsis} available at {\tt https://pypi.org/project/pymsis/}).  We assume a single relative abundance pattern calculated at typical parameters, i.e., noon on 2000-09-15, solar radio flux F10.7 $=$ 100 solar flux unit (SFU, 1~SFU = $10^{-22}$\,W\,m$^{-2}$\,Hz$^{-1}$), geomagnetic index Ap $=$ 5.  Note that N/O ratios are constant to within 10\% below 120\,km for various different input parameters to the NRL model and that a combined N and O density varies within less than 1\% (see below for more details), which is usually less than the statistical errors.  The {\tt cabs} component has a single parameter, i.e., the hydrogen column density, assuming that $N_{\rm H} \times 1.21$ = $N_{\rm e}$ expected for the solar abundance.  This is usually good for cosmic plasmas but is not realistic for the air.  To adapt {\tt cabs} for the air composition, we compute the total number of protons in the air, i.e., $\Sigma Z_{\rm i} N_{\rm i}$ with i representing elements N, O, and Ar, $Z$ being the atomic number, $N$ being the column density.  Then, we take it as an equivalent hydrogen column density for the input of the {\tt cabs} model.  In this calculation, we utilize the abundances of N, O, and Ar obtained from the {\tt vphabs} component, so that the {\tt vphabs} model becomes consistent with the {\tt cabs} model.

\begin{figure}
\includegraphics[width=35pc,angle=0]{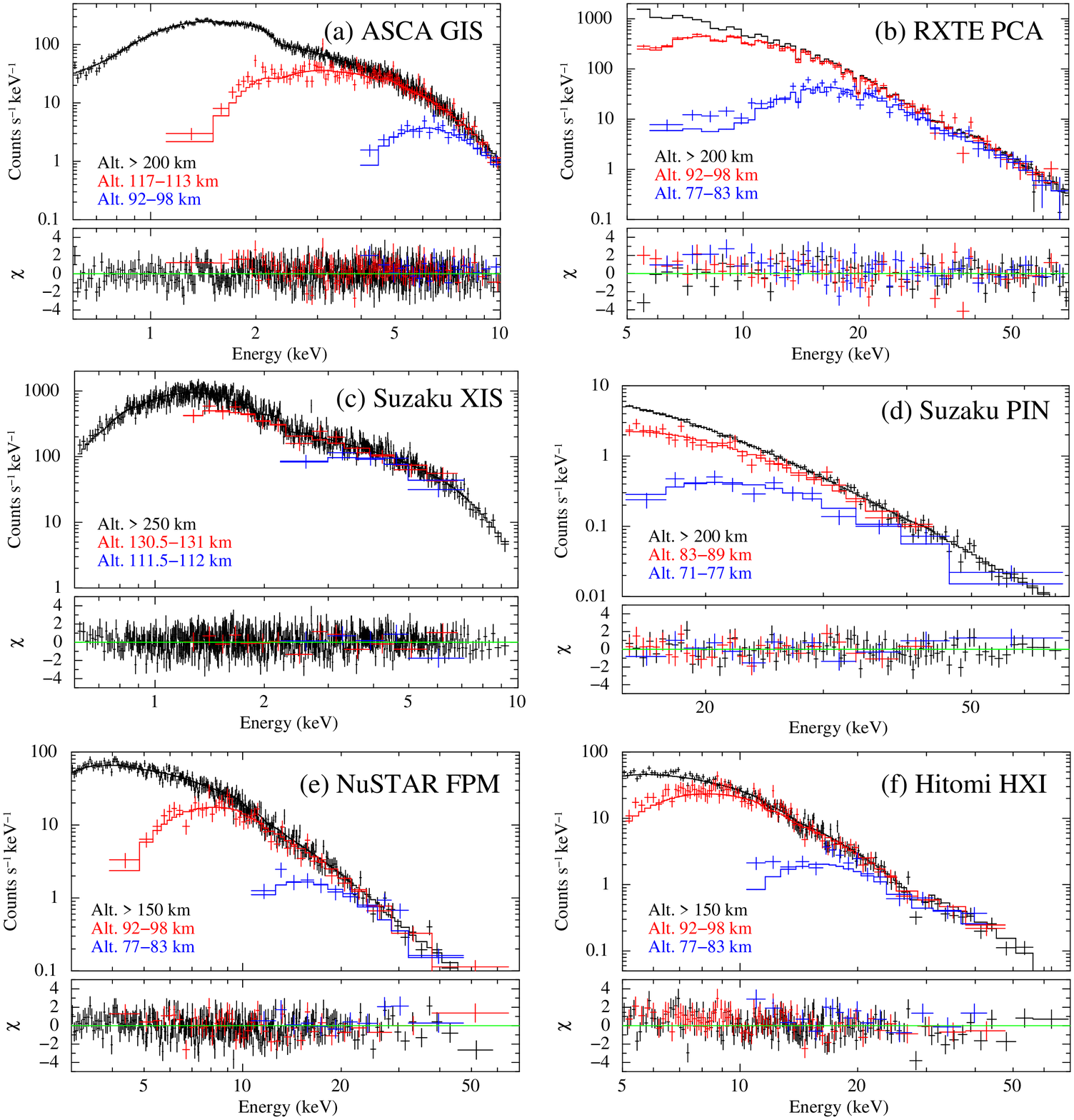}
\caption{(a) Example X-ray spectra obtained with the ASCA/GIS.  The occultations used are the same as Fig.~\ref{fig:lc} (a).  Data in black, red, and blue are taken at different altitude layers.  It is clear that the X-ray intensity decreases from the low-energy band, as the layer goes deeper.  The data in black are fitted with {\tt TBabs} $\times$ {\tt powerlaw}, and those in red and blue are fitted with {\tt vphabs} $\times$ {\tt cabs} $\times$ {\tt TBabs} $\times$ {\tt powerlaw} (see text for details).  Solid lines are the best-fit models.  (b)--(f) Same as (a) but for RXTE/PCA, Suzaku/XIS, Suzaku/PIN, NuSTAR/FPM, and Hitomi/HXI.}
\label{fig:spec}
\end{figure}

The only free parameter in this fitting procedure is the N column density in the atmospheric absorption component {\tt vphabs}.  The best-fit N column density and the assumed N/O ratio allow us to calculate a total N+O (molecules and atoms are combined) column density at each altitude layer.  All the spectra are well reproduced with this model.  Figure~\ref{fig:spec} shows example spectra with the best-fit models for each of the six instruments.

We evaluate the uncertainties on the N(+O) column density caused by the TLE errors estimated in the previous section.  To this end, we reanalyze the six occultations in Obs.ID 70018 obtained with RXTE.  We artificially shift the satellite position by $\pm$1-$\sigma$ in three directions, based on Table~\ref{tab:TLEerror}.  From the shifted satellite positions, we recalculate tangent altitudes for each occultation, and remeasure the vertical N+O column density.  Thus-derived column density ratios between $+1\sigma$ and $-1\sigma$ shifts are summarized in Table~\ref{tab:NOerror}.  The total uncertainties in the three directions amount to $\pm$6\%, which are dominated by longitudinal errors as expected from the TLE errors in Table~\ref{tab:TLEerror}.

The O/N$_2$ relative abundance in the air is known to vary with season, latitude, and geomagnetic activity \cite<e.g.,>{2015E&SS....2....1M}.  Therefore, one may expect variations in the O/N atomic number ratio.  Indeed, according to the NRLMSIS 2.0 model, the variation of O/N is as large as $\pm$10\% at 120\,km altitude, but is almost negligible below 100\,km.  To check the maximum effect on the O/N variation, we artificially modify the O/N ratios by $\pm$10\%, and perform the spectral fitting.  As a result, the combined N$+$O column density in the air vary by only $\pm$1\%.  This is smaller than typical statistical uncertainties.  We note that there is also a possibility of a long-term negative trend in the O/N$_2$ ratio \cite{2009CosRe..47..480P}.  However, it is currently insignificant, and thus we do not consider the effect of long-term variability in this paper.  In any case, the error caused by the incorrect N/O ratio should be smaller than statistical uncertainties, because the magnitude of possible long-term variation is comparable with the short-term variation.

\begin{figure}[htbp]
\includegraphics[width=35pc,angle=0]{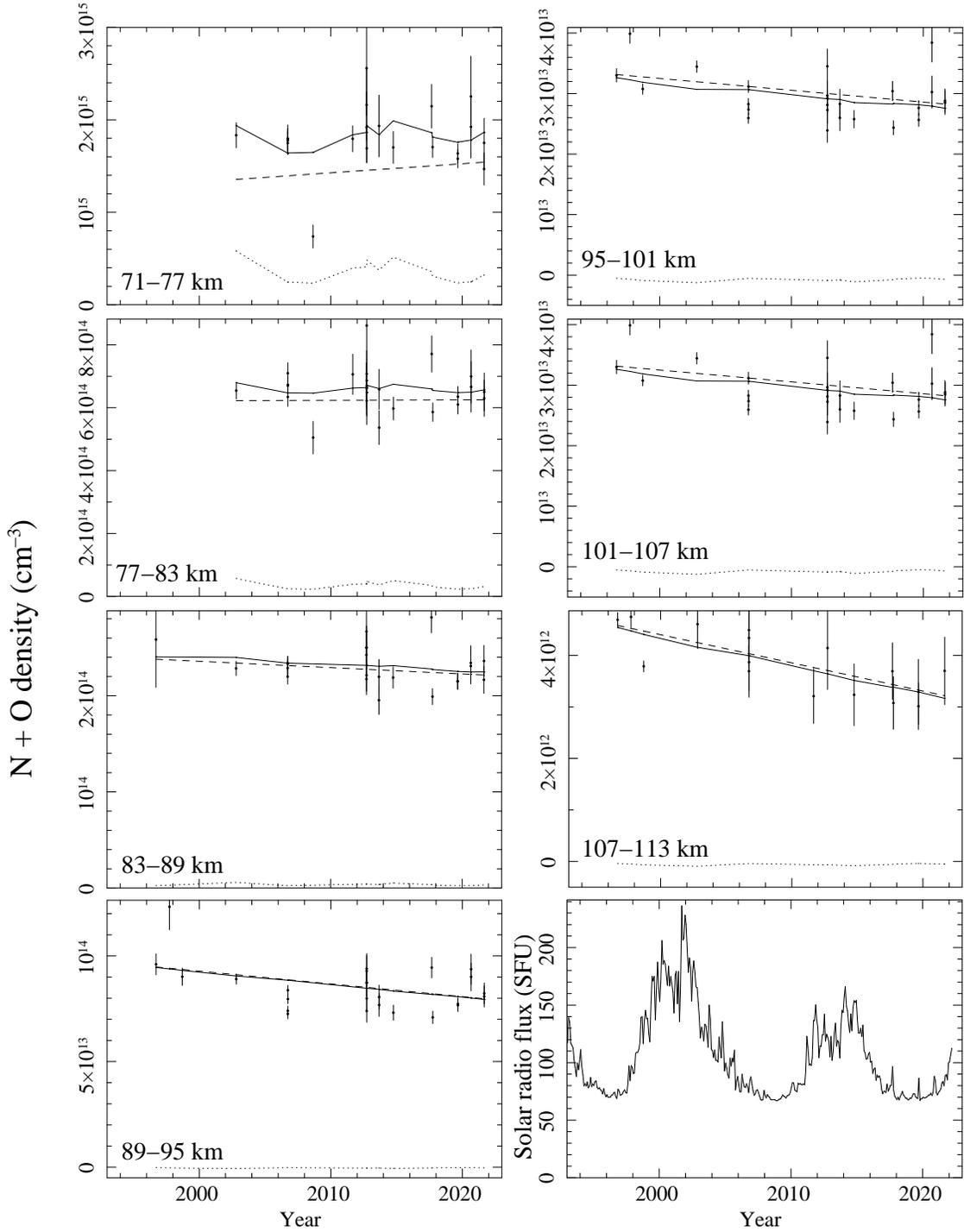}
\caption{Combined N+O density trends at altitude layers indicated in the lower-left corner of each panel.  The plots concentrate on the data taken at Autumn-midnight.  The solid lines represent the best-fit model obtained by the multiple linear regression analysis.  Individual terms, i.e., $a + b \times t$ and $c \times F10.7(t)$ are also shown as dashed and dotted lines, respectively.}  The bottom right panel shows the solar radio (10.7\,cm) flux in solar flux unit (SFU).
\label{fig:density_trend}
\end{figure}

We calculate the local number density at the tangent point by inverting the Abel integral equation (the column density) as we did in Paper~I.  As a result, we obtain vertical density profiles as a function of geometric altitude.  In this inversion process, we arrange the irregular altitude binning for Suzaku/XIS, $\sim$0.2-km bin every $\sim$16\,km, to become uniform 6-km bins.  Then, we generate time series of the N+O density in the time period of 1994--2022 at each altitude.  Figure~\ref{fig:density_trend} shows combined N+O density trends at each of altitude layers, where we concentrate on the data taken during Autumn-midnight.  Note that the temporal coverages by the data are not exactly the same among the plots.  This is partly because the sensitive altitudes are different from instrument to instrument (i.e., ASCA/GSI: 83--116\,km; NuSTAR/FPM: 71--116\,km; Suzaku/XIS: 101--116\,km; Suzaku/PIN: 71--95\,km; NuSTAR/FPM: 71--116\,km; Hitomi/HXI: 71--116\,km as shown in Fig.~\ref{fig:lc}), and partly because we discard unreliable data with statistical errors of $>$20\% as they do not help constrain long-term trends.

We then determine long-term trends and solar responses at all altitude layers, by using a multiple linear regression (MLR) model to the N+O density time series.  To this end, we utilize a python package {\tt sklearn.linear$\_$model.LinearRegression developed by \citeA{JMLR:v12:pedregosa11a}.}  Because we focus on two months around spring and autumn equinoxes, our data are inevitably de-seasonalized if we perform the MLR fitting for the two seasons separately.  The effects of tidal waves (diurnal variations) can be also suppressed by analyzing the data taken around noon and midnight, separately (see Fig.~\ref{fig:obs_loc} and Table~\ref{tab:data}).  Thus, we have four time series of the air density at each altitude.  The MLR model is expressed as:
\[
n_{\rm MLR}(t) = a + b \times t + c \times F10.7(t),
\]
where $n_{\rm MLR}(t)$ is the N+O density, $t$ is the time (year) when the data were taken, and $F10.7(t)$ is the solar radio flux which is daily averaged at each occultation date; the solar flux data are taken from {\tt https://lasp.colorado.edu/lisird/data/noaa\_radio\_flux} and {\tt https://www.spaceweather.gc.ca/forecast-prevision/solar-solaire/solarflux/sx-5-en.php} for old and recent data, respectively.   The bottom panel in Fig.~\ref{fig:density_trend} shows the solar radio (10.7\,cm) flux in solar flux unit (SFU).  The constant $a$, long-term trend $b$, and the amplitude to the solar response $c$ are obtained through least square fitting.

The best-fit long-term trends are summarized in Table~\ref{tab:density_trend}.  The errors quoted are one sigma statistical uncertainties, based on the assumption of independent and identically distributed errors in the fitted data (practically, these errors are the ``standard error" returned by the python tool).  It should be noted that some time series are too sparse (number of data points $<$ 4) to derive robust (or unbiased) trends from the MLR fitting, for which we assign ``not available (N.A.)".  In Fig.~\ref{fig:density_trend}, we plot example best-fit models and the data.  The models in this figure are shown for the total (solid) and individual terms, i.e., $a + b \times t$ (dashed) and $c \times F10.7(t)$ (dotted).  The density trends obtained are generally negative, which is at least qualitatively consistent with past observations \cite{2014GeoRL..41.6919S,2021JASTP.22005650B} and numerical simulations \cite{2006JASTP..68.1879A,2019JGRA..124.3799S}.  There are no obvious differences among the different seasons and local times.  Therefore, we calculated error-weighted means at all altitude layers.  Figure~\ref{fig:trend_amp} left and right show the mean values of the density trends and solar response terms, respectively.  The long-term trend is about $-5$\%/decade throughout the altitudes investigated, except for $\sim$105\,km altitude where the trend shows a local minimum of $-12$\%/decade.  On the other hand, the solar response term is positive everywhere at about 5\%/100 SFU.  The positive response to F10.7 is consistent with past observations of the temperature trends in the MLT \cite<e.g.,>{2016JGRA..121.8951Z,2021JASTP.22005650B}.  

To check the goodness of MLR fittings, we calculate $\chi^2$ and reduced-$\chi^2$ values, where $\chi^2 = \sum_{i=1}^n \frac{(Data_{\rm i} - Model_{\rm i})^2}{\sigma_{\rm i}^2}$, with $i$ being the epoch, $n$ being the number of data points, and $\sigma$ being one sigma error, and the reduced-$\chi^2$ values are defined as $\chi^2$ divided by the degree of freedom which is the number of data points minus number of free parameters.  If the reduced-$\chi^2$ becomes about unity, then we cannot rule out the model (or we can consider that the model is appropriate).  In this process, we add TLE-induced 1-$\sigma$ errors on the N+O density, i.e., $\pm6$\% as shown in Table~\ref{tab:NOerror}.  Because the TLE-based satellite's positions are used only for ASCA and NuSTAR, we introduce this error only for data taken with the two satellites.  Table~\ref{tab:chi2} lists the reduced $\chi^2$ values, from which we can see that the fit quality is mostly satisfactory.  

However, there are several poor quality fits, too.  They might be improved by introducing additional terms in the MLR model.  In fact, \citeA{2017JASTP.163....2L} summarized several possible drivers of long-term trends in addition to the greenhouse gases.  These include secular changes in the geomagnetic activity (Ap) as well as dynamical variations such as Quasi-Biennial Oscillation (QBO), El Nino Southern Oscillation (ENSO), and Arctic Oscillation (AO).  In addition, long-term trends in the atmospheric circulation and wave activity may be also important \cite{2019AnGeo..37..851W}.  Although their effects are not well known, \citeA{2021JASTP.22005650B} tried to incorporate those effects by introducing inter-hemispheric and intra-hemispheric indices calculated from temperature data provided by the National Center for Environmental Prediction/National Center for Atmospheric Research.

Therefore, we perform the MLR fitting by introducing additional terms of Ap ({\tt https://www.gfz-potsdam.de/kp-index}), QBO ({\tt https://iridl.ldeo.columbia.edu/SOURCES/.Indices/.QBO/.QBO\_realtime/}), ENSO ({\tt https://psl.noaa.gov/enso/mei/}), AO ({\tt https://www.daculaweather.com/4\_ao\_index.php}), inter-hemispheric, and intra-hemispheric coupling indices.  We take the inter- and intra-hemispheric coupling indices from \citeA{2021JASTP.22005650B}.  To see their individual contributions, we perform the MLR fitting by adding each of Ap, QBO, ENSO, AO, and a combination of inter- and intra-hemispheric terms to the original model, i.e., linear and solar response terms.  Note that the intra-hemispheric terms in \citeA{2021JASTP.22005650B} include effects by QBO and ENSO.  Therefore, when adding the combination of the inter- and intra-hemispheric terms, we effectively take account of several effects at the same time.  We realized that none of these additional parameters improve the fit quality.   In some cases, the reduced-$\chi^2$ becomes even worse due to the increased number of free parameters.  Moreover, we check that the long-term trends and the magnitudes of the solar response remain consistent within the errors before and after introducing the additional terms, which assures the robustness of the result shown in Fig.~\ref{fig:trend_amp}.  Thus, we conclude that the additional indices do not play an important role in controlling the long-term variation of the neutral density in the MLT region.  

\begin{figure}[htbp]
\includegraphics[width=33pc,angle=0]{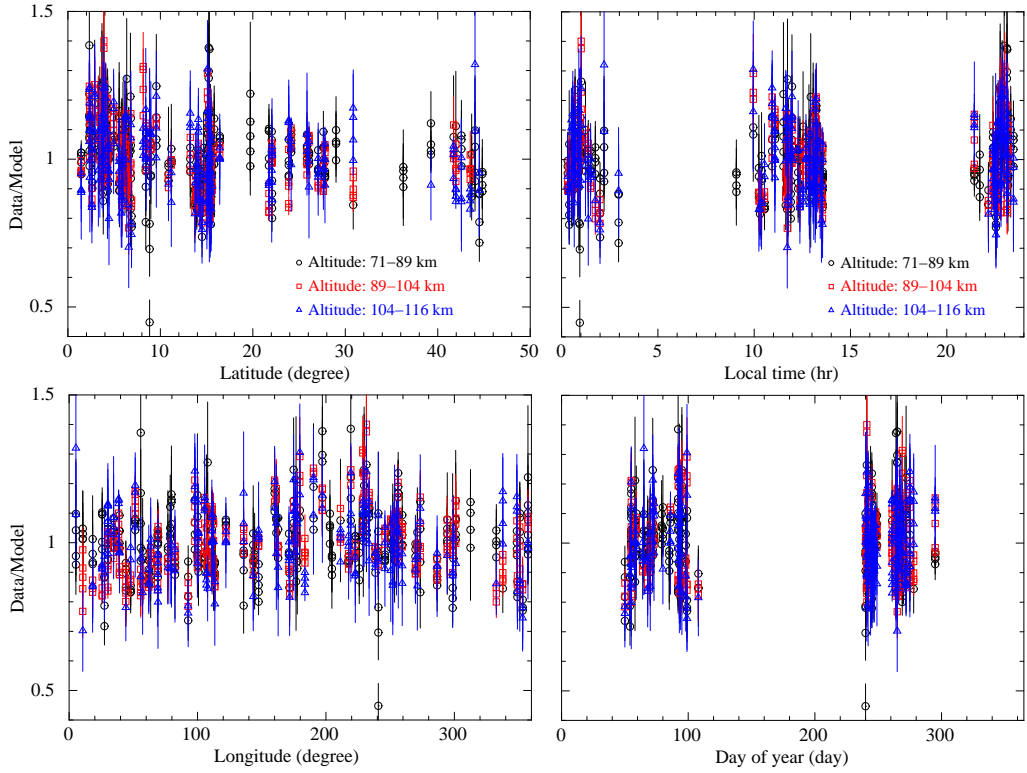}
\caption{ Upper left: Ratios of data to the best-fit MLR model (as shown in Fig.~\ref{fig:density_trend}) as a function of the latitude.  Black open circles, red open boxes, and blue open triangles are obtained at tangent altitude ranges of 71--89\,km, 89--104\,km, and 104--116\,km, respectively.  Lower left, upper right, lower right: Same as upper left, but the x-axes are the longitude, the local time, and the day of year, respectively.}
\label{fig:residual}
\end{figure}

\begin{figure}[htbp]
\includegraphics[width=30pc,angle=0]{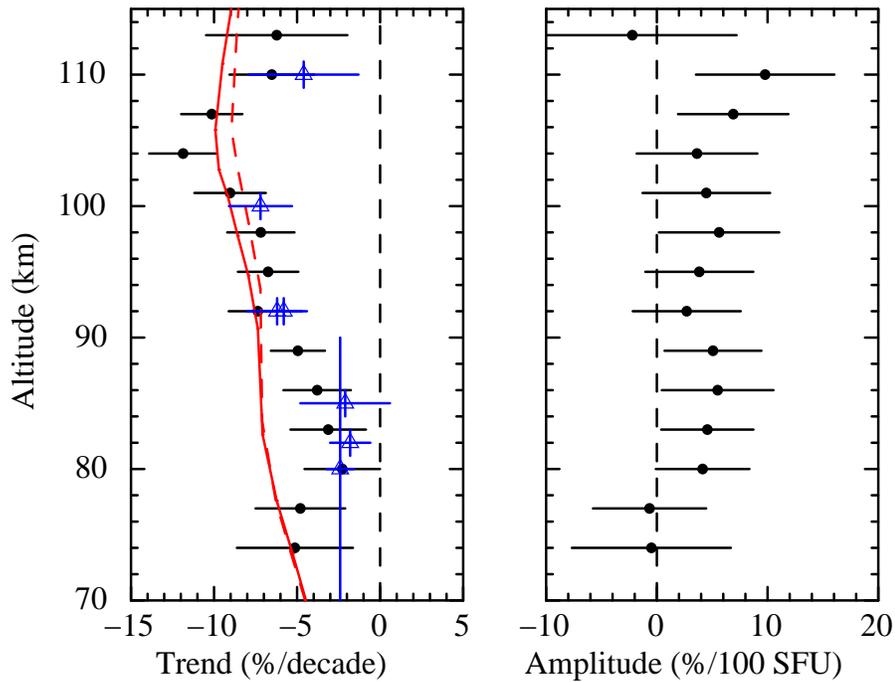}
\caption{Left: Linear trends in N+O density as a function of geometric altitude.   The solid and dashed lines represent numerical simulations computed for the solar maximum and minimum conditions, respectively \cite{2019JGRA..124.3799S}.  The triangle data in blue show density trends inferred from past measurements of atmospheric falling.  Right: N+O density amplitude to the radio solar flux as a function of geometric altitude.  }
\label{fig:trend_amp}
\end{figure}

As another effort to interpret the deviation from the long-term density trend, we also search for possible relationships between the residuals and physical parameters.  Figure~\ref{fig:residual} presents data-to-model ratios against such parameters as the latitude, longitude, local time, and day of year, from which we cannot find meaningful patterns.

There are several possibilities to explain the bad fits.  One obvious cause is insufficient knowledge of atmospheric wave activities.  For example, anomalous short-period gravity waves originating from the lower atmosphere could easily produce outliers in the density time series \cite<e.g.,>{2018JAtS...75.3613S,2018JAtS...75.3635Y}.  In addition, short-term variations owing to planetary waves, which are often removed by averaging a number of data in a month or two, are difficult to be suppressed in our sparse data sets; we have only a few data groups in one season.   In particular, the semidiurnal tide shows a clear solar cycle dependence \cite{2019AnGeo..37..851W}.  Another possibility associated with planetary waves is a stratospheric sudden warming (SSW), which is a sudden temperature increase by more than a few dozen degrees near north pole, caused by a rapid amplification of planetary waves propagating upward from the troposphere \cite<e.g.,>{1971JAtS...28.1479M,2012JGRD..11716101T}.  SSW signatures are very small for the latitude band considered here, because it is a primarily polar phenomenon.  However, we may not fully rule out the possibility that our data are affected by SSWs.  Given that SSWs occur typically in late December to January and early February, we searched for signatures in our data taken in spring.  After all, we found no significant density differences between SSW and non-SSW springs, suggesting that SSWs do not play important roles in our data.  Another possible contributor to the bad fits is climatological geophysical variations within the seasonal, local time, latitudinal bins.  Such effects could be mitigated by analyzing the data divided by the appropriate model such as MSIS, which will be accounted for in the future work.  We should point out however, that as shown in Fig.~\ref{fig:residual}, the data/model ratios do not show clear systematic relationships with any physical parameters examined.  This fact suggests that normalization by the MSIS model can not help improve the MLR fitting quality.  Finally, we note that the TLE error is another source that could produce outliers for ASCA and NuSTAR.

\section{Discussion}

We obtained long-term trends of total number densities of N and O (both atoms and molecules are combined) in the MLT region, based on atmospheric occultations of the Crab Nebula observed with X-ray astronomy satellites, ASCA, RXTE, Suzaku, NuSTAR, and Hitomi.  In this section, we compare our results with past observations and numerical simulations.  

As we described above, direct density measurements in the MLT region, in particular long-term trends in density, is scarce.  As far as we know, the only long-term density trend has been measured for electron densities, which have been measured with sounding rockets for more than 80\,years, providing us with long-term trends \cite<e.g.,>{2017JASTP.163...78F}.  In general, an increase in the electron density of the order of $\sim$1\%/yr was found in the MLT region.  This, combined with the vertical increase upward in the electron density, suggests a downward shifts of the electron density profile.  This is qualitatively consistent with the atmospheric contraction/descent due to the greenhouse cooling.  However, quantitative evaluation of the trend in total density has not yet been obtained.  

Except for the electron density, density trends in the MLT region have been inferred from measurements of atmospheric falling from a variety of methods.  These include (1) the height of the $E$-layer's peak electron density \cite<e.g.,>{2008AnGeo..26.1189B}, (2) the mesopause height, i.e., the height at the local temperature minimum \cite<e.g.,>{2019JGRD..124.5970Y}, (3) the centroid height of the Na layer \cite<e.g.,>{1997JASTP..59.1673C}, (4) the meteor peak flux altitude \cite<e.g.,>{2014GeoRL..41.6919S}, (5) the ionospheric reflection height in the low-frequency (LF) range \cite<e.g.,>{2015AdSpR..55.1764P,2017JASTP.163...23P}, (6) the pressure altitude measured by atmospheric occultations and Earth's limb emission \cite<e.g.,>{2021JASTP.22005650B}.  In addition, there are other indirect observables to estimate the atmospheric falling, including the height of the turbopause \cite{2009CosRe..47..480P,2016ACP....16.2299H} and the height of noctilucent clouds, also known as polar mesospheric clouds \cite<e.g.,>{2017JASTP.162...79F,2021JASTP.21405378L}.  In this paper, we do not discuss them, because the turbopause heights are known to be problematic as discussed in \citeA{2015JGRA..120.2347L,2017JASTP.163....2L} and the altitudes of noctilucent clouds depend not only on the density profile but also strongly depend on the temperature and the water vapor concentration in the atmosphere.  Furthermore, there is evidence that space traffic has a strong influence on the interannual variability of the bright noctilucent clouds \cite{Stevens2022}.

The air density corresponds one-to-one to the altitude, given the locally hydrostatic balanced relationship: $d\rho \propto -\rho~dh$.  Therefore, we can convert the altitude to the density by using some atmospheric models.  Consequently, we can convert the speed of atmospheric vertical shifts to the rate of density change.  According to the NRLMSIS 2.0 model \cite{2021E&SS....801321E}, the air density decreases with increasing altitude at a rate of 16$\pm$1\%/km in the MLT region.  To check possible systematic uncertainties on density models, we also examined another empirical atmospheric density model, Jacchia-Bowman 2008 \cite{Bowman2008}.  This model gives a density gradient of $\sim$17\%/km at 90\,km altitude, which is consistent with the NRLMSIS 2.0 model.  Thus, we multiply an altitude trend by a value of 16$\pm$1\%/km to derive a density trend.  The result is summarized in Table~\ref{tab:altitude_trend} and Fig.~\ref{fig:density_trend}.  We note that \citeA{2014GeoRL..41.6919S} showed the density trend inferred from the meteor peak flux altitude, but did not explicitly give the altitude trend.  Therefore, we calculated the altitude trend based on Fig.~4 in \citeA{2014GeoRL..41.6919S}, by performing the same MLR analysis as we did in the previous section.  We can see that our measurements are generally consistent with past observations at every altitude.  

It should be noted, however, that our trend slopes in density are systematically steeper (albeit within error bars) than those inferred from past observations.   Resolving this issue is beyond the scope of this paper, but one problem is that our conversion process from altitude trends to density trends for past observations may be too simple.   We also note that the vertical trend profile exhibits a local minimum at $\sim$105\,km, with a decline rate of -12\%/decade.  In fact, past observations show the largest (negative maximum) trend near 100\,km, which is at least qualitatively consistent with our measurement.  Unfortunately, past observations do not cover the altitude range of 100--110\,km.  Therefore, revealing the density trend in this altitude range is left as an important future work.

The long-term (2--3 decades) density decrease can be also inferred from model comparisons between NRLMSISE-00 and NRLMSIS 2.0, because the former model is based primarily on mass spectrometer data that are now 35--50 years old, whereas the latter one assimilates extensive new (since 2000) measurements and analyses of temperature in the mesosphere, stratosphere, and troposphere, as well as many years of new atomic O and H measurements in the mesosphere.  From Fig.~18 (c) in \citeA{2021E&SS....801321E}, MSIS 2.0 N$_2$ densities near 100\,km and above are $\sim$20\% lower than MSISE-00.  This is in reasonable agreement with our result.

As shown in Fig.~\ref{fig:trend_amp} left, our results are also in reasonable agreement with the state-of-the-art numerical simulations using WACCM‐eXtended, in which trends are calculated from 5-year simulations for the years 1972--1976 and 2001--2005, which serves as a small ensemble of similar years and are averaged over each of the two ensembles \cite{2019JGRA..124.3799S}.  This model takes into account all key trace constituents, i.e., CO$_2$, CH$_4$ (a precursor of H$_2$O), H$_2$O, and O$_3$.  The model was computed for the solar maximum and minimum conditions, the results of which are illustrated in Fig.~\ref{fig:trend_amp} left as the solid and dashed lines, respectively.   We can see that the difference between the solar minimum and maximum conditions are small in the region of interest.  It is interesting to note that numerical experiments exhibit a local maximum of a density decline just above 100\,km (the local minimum is more evident in the original figure, Fig.~3 in \citeA{2019JGRA..124.3799S}, showing a wider altitude range).  \citeA{2006JASTP..68.1879A} first pointed out that this peak is created by the effects of H$_2$O and O$_3$; the cooling effect on the density accumulates with altitude up to $\sim$110\,km where in situ H$_2$O and O$_3$ forcings disappear.  The density trend obtained in this work shows a local (negative) maximum near 105\,km altitude.  This seems to be the first observational evidence for the local maximum in the density decline due to H$_2$O and O$_3$.

There is about 20\,yr time difference between the simulations by \citeA{2019JGRA..124.3799S} and our data sets, which is subject to possible systematic errors, because temporal variations of greenhouse gas concentrations vary from period to period.  The simulations took the increasing rates of CO$_2$, CH$_4$, and chlorofluorocarbon (sensitive to O$_3$) to be $+$4.5\%/decade, $+$6.3\%/decade, and $+$30\%/decade, respectively.  In the middle and upper stratosphere, H$_2$O is mainly formed by CH$_4$ oxidation process.  According to \citeA{2013JGRA..118.3846Q}, the 6.3\% increase in CH$_4$ concentration results in an increase of 2\%/decade in H$_2$O concentration in the upper atmosphere.  This H$_2$O increase seems to be incorporated in the simulations by \citeA{2019JGRA..124.3799S}, although it is difficult to evaluate the H$_2$O increasing rate quantitatively from their paper alone.  

Realistic trends of greenhouse gases in the time period of our data (1994--2022) would be different from the values assumed in the simulations by \citeA{2019JGRA..124.3799S}.  In particular, the O$_3$ trend changed from negative (ozone depletion) to none or positive around 1990 as a consequence of the Montreal protocol in 1987.  Because O$_3$ is a strong absorber of the solar ultra-violet radiation, the temperature (and density) response to O$_3$ increase is opposite to that of the other major greenhouse gases.  The O$_3$ increase has a heating effect on the MLT region via reduced solar heating, which was demonstrated by numerical simulations by \citeA{2013JGRD..11813347L}.  Given that O$_3$ is increasing in the recent three decades \cite{acp-20-8453-2020,2020Atmos..11..795B}, O$_3$ should work to expand the middle and upper atmosphere during the time period of our data sets.  Therefore, a proper incorporation of the effect of O$_3$ in the numerical simulations would make the trend slope in density larger (closer to zero) to some extent.

As for the increasing rate of H$_2$O concentration, recent measurements of water vapor with the Sounding of the Atmosphere using Broadband Emission Radiometry (SABER) and the Aura Microwave Limb Sounder found a global increase of 3$\pm$1.5\%/decade near 80\,km altitude \cite{2019GeoRL..4613452Y}.  This is the same level as that assumed in the simulations (likely 2\%/decade).  Therefore, the simulation is realistic from the point of view of the H$_2$O concentration.

The CO$_2$ concentration in the upper atmosphere has been debated \cite[for a review]{2019JASTP.189...80L}.  However, the most recent result by \citeA{2018JGRA..123.7958R} who re-analyzed SABER data between 2002 and 2016 came to a conclusion that CO$_2$ density trends below 90\,km are consistent with the tropospheric value of 5.5\%/decade, whereas above 90\,km the trend becomes higher, reaching a maximum value of 8.2\%/decade at $\sim$105\,km altitude for bimonthly time bins (or 10\%/decade for monthly time bins).  This is consistent with earlier results obtained from the solar occultation data with the Atmospheric Chemistry Experiment Fourier Transform Spectrometer \cite{2012NatGe...5..868E}.  In this context, it seems that the most reliable trend in the MLT is twice larger than the value used in the simulations by \citeA{2019JGRA..124.3799S}.  

In this way, all of these input parameters including CO$_2$, CH$_4$, H$_2$O, and O$_3$ concentrations must be updated to be consistent with the observational data.  Because the density response to the increasing greenhouse gases is nonlinear, it is necessary to perform numerical simulations that use physical parameters consistent with observational data.

\section{Conclusions}

We obtained long-term density trends in the last 28\,yr (1994--2022) of the Earth's upper atmosphere at geometric altitudes between 71 and 116\,km with a resolution of 3\,km, based on atmospheric occultations of the Crab Nebula observed with X-ray astronomy satellites, ASCA, RXTE, Suzaku, NuSTAR, and Hitomi.  The data taken in different seasons (spring and autumn) and different local times (noon and midnight) are combined to provide a single vertical profile of trend slopes.  The density trends are overall negative at roughly $-$5\%/decade.   This is roughly consistent with inferences from past measurements of the rates of settling atmosphere in the MLT region.  In the 100-110\,km altitude, we found an exceptionally high density decline of about $-$12\%/decade.  We believe that this peak is the first observational evidence for strong cooling due to water vapor and ozone near 110\,km, which was first identified in a numerical simulation by \citeA{2006JASTP..68.1879A}.  However, this feature needs to be verified by further observations as well as numerical simulations, which use physical parameters that correspond to the time periods of observations.

\acknowledgments
All the data used in this paper can be found at NASA's HEASARC website, {\tt https://heasarc.gsfc.nasa.gov/cgi-bin/W3Browse/w3browse.pl}.  We thank Mina Ogawa and Tadayasu Dotani for providing us with information about position uncertainties on Japanese X-ray astronomy satellites, and Ryo Iizuka for suggesting a method to correct for Hitomi satellite's positions in the housekeeping file.  Craig B.\ Markwardt and Wataru Iwakiri gave us helpful comments on the analysis of RXTE data.  Brandon Rhodes helped us use {\tt skyfield} to calculate satellites' positions in various coordinates, using TLE$+$SGP4.  T.S.\ Kelso pointed out bad TLE data for the ASCA satellite.  We also thank Ryosuke Yasui, Hitoshi Fujiwara, and Titus Yuan for their expertized comments on the upper atmosphere.  We are grateful to two referees who gave us constructive comments that improved the quality of the paper.  This work was supported by the Japan Society for the Promotion of Science KAKENHI grant numbers 20K20935 (SK and MST).  This work was partly supported by Leading Initiative for Excellent Young Researchers, MEXT, Japan.


\clearpage

\begin{table}
 \caption{Capabilities of X-ray instruments used in this work}
 \label{tab:instruments}
\begin{tabular}{lcccccc}
\hline
Parameter & ASCA/GIS  & RXTE/PCA & Suzaku/XIS & Suzaku/PIN & NuSTAR/FPM & Hitomi/HXI \\
\hline
Field of view & 24$^\prime$ in radius & 1$^\circ$ FWHM & 18$^\prime$$\times$18$^\prime$ & 34$^\prime$$\times$34$^\prime$  & 10$^\prime$$\times$10$^\prime$ & 9$^\prime$$\times$9$^\prime$  \\
Spatial resolution$^a$ & 3$^\prime$ & --- & 2$^\prime$ & --- & 1$^\prime$ & 1$^\prime$.7 \\
Energy resolution$^b$ & 13 at 6\,keV & 5 at 6\,keV & 46 at 6\,keV & 7 at 20\,keV & 45 at 20\,keV & 14 at 14\,keV \\
Time resolution$^c$ & 1\,$\mu$s & 0.24\,ms / 4\,s & 0.1\,s & 0.1\,ms & 61\,$\mu$s & 26\,$\mu$s  \\
\hline
\multicolumn{7}{l}{\small $^a$The angular resolution is given by half power diameter.  RXTE/PCA and Suzaku/PIN are non-imaging instruments.}\\
\multicolumn{7}{l}{\small $^b$The energy resolution is given by $E/\Delta E$.}\\
\multicolumn{7}{l}{\small $^c$The RXTE/PCA time resolutions of 0.24\,ms and 4\,s are for ObsIDs 70018 and 92018, respectively.}\\
\end{tabular}
\end{table}

\begin{table}
 \caption{TLE errors for the RXTE satellite}
 \label{tab:TLEerror}
\begin{tabular}{lcc}
\hline
Direction & 1-$\sigma$ C.L. (km)  & 90\,\% C.L. (km)\\
\hline
Longitude & $\pm1.1$ & $\pm1.4$\\
Latitude & $\pm0.22$ & $\pm0.35$\\
Altitude & $\pm0.13$ & $\pm0.21$\\
\hline
\end{tabular}
\end{table}

\begin{table}
 \caption{N+O column density errors caused by 1-$\sigma$ TLE errors}
 \label{tab:NOerror}
\begin{tabular}{lcccccc}
\hline
Data group & \multicolumn{3}{c}{Ratios of N$+$O Column Densities}  & Total difference \\
& Longitude & Latitude & Altitude & \\
& $\frac{N_{\rm A,N+O}({\rm Sat.pos.}-1.1\,{\rm km})}{N_{\rm A,N+O}({\rm Sat.pos.}+1.1\,{\rm km})}$  & $\frac{N_{\rm A,N+O}({\rm Sat.pos.}-0.17\,{\rm km})}{N_{\rm A,N+O}({\rm Sat.pos.}+0.17\,{\rm km})}$ & $\frac{N_{\rm A,N+O}({\rm Sat.pos.}-0.1\,{\rm km})}{N_{\rm A,N+O}({\rm Sat.pos.}+0.1\,{\rm km})}$ & \\
\hline
7 & 1.12 & 0.98 & 0.96 & $\pm$6\%\\
8 & 1.13 & 0.99 & 0.96 & $\pm$7\% \\
9 & 1.11 & 0.99 & 0.96 & $\pm$6\% \\
10 & 0.89 & 0.99 & 0.97 & $\pm$6\% \\
11 & 0.89 & 0.99 & 0.96 & $\pm$6\% \\
12 & 0.88 & 0.99 & 0.95 & $\pm$7\% \\
\hline
\end{tabular}
\end{table}

\begin{table}
 \caption{Density trend at each altitude layer}
 \label{tab:density_trend}
\begin{tabular}{lcccccc}
\hline
Altitude & Autumn-noon& Autumn-midnight  & Spring-noon & Spring-midnight & Mean\\
 (km) & (\%/decade) & (\%/decade) & (\%/decade) & (\%/decade) & (\%/decade) \\
\hline
71--77 & $-8.5\pm5.2$ & $+5.5\pm7.6$ & $-8.2\pm8.1$&  $-5.9\pm9.0$ & $-5.1\pm3.5$\\
74--80 & $-7.2\pm4.1$ & $-0.1\pm5.5$ & $-6.7\pm6.6$ & $-3.6\pm6.7$ & $-4.8\pm2.7$\\
77--83 & $-4.0\pm3.5$ & $+0.2\pm4.3$ & $+0.8\pm6.1$ & $-4.7\pm5.5$ & $-2.3\pm2.3$\\
80--86 & $-3.8\pm3.7$ & $-2.7\pm4.0$ & $+3.9\pm7.3$ & $-5.9\pm5.1$ & $-3.1\pm2.3$\\
83--89 & $-4.6\pm2.8$ & $-2.9\pm3.0$ & N.A. & N.A. & $-3.8\pm2.0$\\
86--92 & $-6.1\pm2.0$ & $-3.0\pm2.7$ & N.A. & N.A. & $-4.9\pm1.6$\\
89--95 & $-7.4\pm2.0$ & $-7.1\pm3.5$ & N.A. & N.A. & $-7.4\pm1.8$\\
92--98 & $-7.8\pm2.2$ & $-5.7\pm3.5$ & $+1.1\pm8.5$ & $-3.9\pm13.9$ & $-6.7\pm1.8$\\
95--101 & $-7.9\pm2.6$ & $-6.6\pm3.6$ & $-1.0\pm9.9$ & $-6.9\pm15.9$  & $-7.2\pm2.0$\\
98--104 & $-8.3\pm2.9$ & $-10.2\pm3.4$ & $-4.6\pm10.9$ & $-12.7\pm16.7$ & $-9.0\pm2.2$\\
101--107 & $-9.3\pm3.4$ & $-14.0\pm2.7$ & $-6.5\pm10.9$ & $-1.3\pm17.6$ & $-11.9\pm2.0$\\
104--110 & $-6.3\pm3.0$ & $-13.4\pm2.4$ & $+2.6\pm10.0$ & $-11.3\pm16.9$ & $-10.1\pm1.8$\\
107--113 & $+2.0\pm4.1$ & $-14.2\pm3.4$ & N.A. & $+23.1\pm14.0$ & $-6.5\pm2.6$\\
110--116 & $+4.2\pm5.5$ & $-21.3\pm6.7$ & N.A. & N.A. & $-6.2\pm4.3$\\
\hline
\multicolumn{6}{l}{\small N.A. stands for not available.}\\
\end{tabular}
\end{table}

\begin{table}
 \caption{Goodness of MLR fittings}
 \label{tab:chi2}
\begin{tabular}{lcccccc}
\hline
Altitude (km) & \multicolumn{4}{c}{Reduced chi-squares} \\
& Autumn-noon  & Autumn-midnight & Spring-noon  & Spring-midnight \\
\hline
71--77   & 1.705 & 3.842 & 0.939 & 2.364 \\
74--80   & 1.296 & 1.157 & 0.844 & 0.879 \\
77--83   & 1.352 & 1.088 & 1.226 & 0.538 \\
80--86   & 1.493 & 0.967 & 1.433 & 0.544 \\
83--89   & 1.277 & 0.817  & N.A. & N.A. \\
86--92   & 1.131 & 1.417 & N.A. & N.A. \\
89--95   & 1.618 & 5.682 & N.A. & N.A. \\
92--98   & 1.699 & 5.710 & 2.719 & 2.351 \\
95--101  & 1.311 & 3.729 & 3.112 & 2.582 \\
98--104  & 1.228 & 1.956 & 2.352 & 1.745 \\
101--107 & 1.083 & 1.075 & 1.168 & 1.293 \\
104--110 & 0.745 & 0.598 & 0.610 & 0.683 \\
107--113 & 0.755 & 0.647 & N.A. & 0.488 \\
110--116 & 0.641 & 1.367 & N.A. & N.A. \\
\hline
\multicolumn{5}{l}{\small N.A. stands for not available.}\\
\end{tabular}
\end{table}

\begin{table}
 \caption{Altitude $\rightarrow$ density trends in the MLT region}
 \label{tab:altitude_trend}
\begin{tabular}{lcccccc}
\hline
Target & Altitude & Obs.\ Period & Altitude trend & Density trend$^a$ & Reference\\
(Method) & (km) & (yr)  &  (km/decade) & (\%/decade) & \\
\hline
$E$-layer $n_{\rm e}$ peak height &  110 & 1957--2005 & $-0.029\pm$0.020 &  $-4.6\pm3.2\pm0.03$ & \citeA{2008AnGeo..26.1189B} \\
(Ionosonde) & & & & & \\
High mesopause height & 100 & 1990--2018 & $-0.45\pm$0.09 & $-7.2\pm1.4\pm0.5$ & \citeA{2019JGRD..124.5970Y}\\
(Na lidar) & & & & & \\
Meteor peak flux altitude & 92 & 2002--2014 & $-0.58\pm$0.01  & $-5.8\pm$1.1 & \citeA{2014GeoRL..41.6919S} \\
(Specular meteor echo) & & & & & \\
Na layer height & 92 & 1972--1994 & $-0.37\pm$0.09 & $-6.2\pm1.4\pm0.4$ & \citeA{1997JASTP..59.1673C} \\
(Na lidar) & & & & & \\
Low mesopause height &  85 & 1990--2018 & $-0.13\pm$0.16 & $-2.1\pm2.6\pm0.1$ & \citeA{2019JGRD..124.5970Y}\\
(Na lidar) & & & & & \\
Radio reflection height & 82 & 1959--2009 & $-0.114\pm$0.078 & $-1.8\pm1.2\pm0.01$ & \citeA{2017JASTP.163...23P} \\
(Phase-height experiment) & & & & & \\
Pressure altitude & 50--91 & 1991--2020 & $-0.15\pm$0.05 & $-2.4\pm0.8\pm0.02$ & \citeA{2021JASTP.22005650B}\\ 
(Satellites' remote sensing) & & & & & \\
\hline
\multicolumn{6}{l}{\small $^a$The first and second error-terms represent intrinsic measurement errors and possible errors on density gradients}\\
\multicolumn{6}{l}{\small ($\pm$1\%/km) in the empirical models, respectively.}\\
\end{tabular}
\end{table}


\clearpage
\appendix
\section{Summary of Atmospheric Occultations Analyzed}

All the data analyzed in this paper are listed in Table~\ref{tab:data}.  For ASCA and NuSTAR, we recalculated satellites' positions to improve the accuracy, using the {\tt skyfield} package into which we gave orbital elements taken from NORAD TLEs ({\tt http://celestrak.com/NORAD/archives/request.php}) with the SGP4 propagator algorithm.  We use original positions for the other satellites, which are provided by the teams.  These satellite positions are the most important ingredients to determine tangent altitudes and tangent points.  We collected data taken in spring ($\pm$30 days around the equinox) and autumn ($\pm$30 days around the equinox).  We also limit the data taken in the northern hemisphere.  The last column denotes the data group.  We merge the data in the same group for our spectral analysis.  


\begin{longtable}[c]{lccccc}
 \caption{Atmospheric occultations of the Crab Nebula analyzed in this paper.}
 \label{tab:data}
 \\
 \hline 
Instrument & UT$^a$ & Tangent point$^b$ & Local time$^c$ & Type & Group\\
(Obs.ID) & (HH:MM:SS) & Long, Lat ($^\circ$) & (HH:MM:SS) & & \\
\hline
 \endfirsthead
 \multicolumn{6}{l}{\small }\\
 \hline
Instrument & UT$^a$ & Tangent point$^b$ & Local time$^c$ & Type & Group \\ 
(Obs.ID) & (HH:MM:SS) & Long, Lat ($^\circ$) & (HH:MM:SS) & & \\
 \hline
 \endhead
 \hline
\multicolumn{5}{l}{\small $^a$Universal time when the line-of-sight tangential altitude becomes 100\,km.}\\
\multicolumn{5}{l}{\small $^b$Latitude and longitude of the tangent point at 100\,km altitude.}\\
\multicolumn{5}{l}{\small $^c$Local time of the tangent point at 100\,km altitude.}\\
 \endfoot 
 \hline
\multicolumn{5}{l}{\small $^a$Universal time when the line-of-sight tangential altitude becomes 100\,km.}\\  
\multicolumn{5}{l}{\small $^b$Latitude and longitude of the tangent point at 100\,km altitude.}\\  
\multicolumn{5}{l}{\small $^c$Local time of the tangent point at 100\,km altitude.}\\
 \endlastfoot
\multirow{5}{*}{
\begin{tabular}{l}
ASCA/GIS\\(10010180)
\end{tabular}}
 & 1994-09-28, 05:24:41 & 106.9204, 41.9572 & 12:32:21 & Setting & 1 \\
 & 1994-09-28, 14:59:35 & 322.4569, 41.5094 & 12:29:24 & Setting & 1 \\
 & 1994-09-29, 00:34:29 & 177.9692, 41.0303 & 12:26:21 & Setting & 1 \\
 & 1994-09-29, 02:10:18 & 153.8861, 40.9471 & 12:25:50 & Setting & 1 \\
 & 1994-09-29, 05:21:56 & 105.7188, 40.7777 & 12:24:47 & Setting & 1 \\
\hline
\multirow{8}{*}{
\begin{tabular}{l}
ASCA/GIS\\(10010190)
\end{tabular}}
 & 1994-10-04, 13:07:13 & 337.7223, 30.8591 & 11:38:05 & Setting & 2 \\
 & 1994-10-04, 22:42:08 & 193.0689, 29.9003 & 11:34:23 & Setting & 2 \\
 & 1994-10-05, 00:17:58 & 168.9556, 29.7475 & 11:33:47 & Setting & 2 \\
 & 1994-10-05, 01:53:47 & 144.8492, 29.5808 & 11:33:10 & Setting & 2 \\
 & 1994-10-05, 03:29:36 & 120.7376, 29.4155 & 11:32:33 & Setting & 2 \\
 & 1994-10-05, 09:52:53 & 24.2916, 28.7546 & 11:30:02 & Setting & 2 \\
 & 1994-10-05, 11:28:42 & 0.1818, 28.5844 & 11:29:25 & Setting & 2 \\
 & 1994-10-05, 13:04:31 & 336.0716, 28.4137 & 11:28:48 & Setting & 2 \\
\hline
\multirow{6}{*}{
\begin{tabular}{l}
ASCA/GIS\\(10403000)
\end{tabular}}
 & 1996-09-17, 08:18:21 & 211.3219, 41.7094 & 22:23:38 & Rising & 3 \\
 & 1996-09-17, 09:23:55 & 38.5445, 6.2673 & 11:58:05 & Setting & 4 \\
 & 1996-09-17, 09:54:07 & 187.4372, 41.5528 & 22:23:51 & Rising & 3 \\
 & 1996-09-17, 10:59:44 & 14.6868, 6.6670 & 11:58:27 & Setting & 4 \\
 & 1996-09-17, 11:29:52 & 163.5178, 41.4414 & 22:23:56 & Rising & 3 \\
 & 1996-09-17, 23:46:12 & 183.7813, 9.6564 & 12:01:18 & Setting & 4 \\
\hline
\multirow{3}{*}{
\begin{tabular}{l}
ASCA/GIS\\(11501000)
\end{tabular}}
 & 1997-09-26, 07:45:40 & 228.6920, 8.1441 & 23:00:25 & Rising & 5 \\
 & 1997-09-26, 09:21:21 & 204.6166, 8.3629 & 22:59:48 & Rising & 5 \\
 & 1997-09-26, 15:44:06 & 108.2979, 9.2428 & 22:57:16 & Rising & 5 \\
\hline
\multirow{5}{*}{
\begin{tabular}{l}
ASCA/GIS\\(11602000)
\end{tabular}}
 & 1998-09-15, 10:12:57 & 183.6161, 43.5114 & 22:27:24 & Rising & 6 \\
 & 1998-09-15, 11:48:28 & 159.6453, 43.5447 & 22:27:02 & Rising & 6 \\
 & 1998-09-15, 13:24:00 & 135.7008, 43.5384 & 22:26:48 & Rising & 6 \\
 & 1998-09-16, 11:41:20 & 160.3667, 43.6404 & 22:22:47 & Rising & 6 \\
 & 1998-09-16, 14:52:23 & 112.4787, 43.6316 & 22:22:17 & Rising & 6 \\
\hline
\multirow{6}{*}{
\begin{tabular}{l}
RXTE/PCA\\(70018)
\end{tabular}}
& 2002-03-05, 14:05:26 & 171.2213, 27.0653 & 01:30:19 & Setting & 7 \\
& 2002-08-29, 22:14:21 & 222.8523, 1.4405 & 13:05:44 & Setting & 8 \\
& 2002-09-26, 02:04:09 & 147.4954, 22.0718 & 11:54:06 & Setting & 9 \\
& 2002-10-22, 14:23:43 & 105.4934, 3.7893 & 21:25:41 & Rising & 10 \\
& 2003-02-27, 04:33:13 & 122.0489, 16.4205 & 12:41:23 & Rising & 11 \\
& 2003-03-27, 03:39:21 & 111.3536, 8.4639 & 11:04:45 & Rising & 12 \\
\hline
\multirow{3}{*}{
\begin{tabular}{l}
RXTE/PCA\\(92018)
\end{tabular}}
& 2006-09-03, 21:09:40 & 245.9062, 27.8003 & 13:33:16 & Setting & 13 \\
& 2006-09-04, 20:43:57 & 250.4656, 26.0700 & 13:25:47 & Setting & 14 \\
& 2006-09-05, 20:18:14 & 254.9570, 24.1404 & 13:18:02 & Setting & 15 \\
& 2006-09-06, 19:52:32 & 259.3940, 22.0419 & 13:10:05 & Setting & 16 \\
& 2006-09-23, 21:05:07 & 25.7248, 21.7392 & 22:48:00 & Rising & 17 \\
& 2006-09-24, 20:39:22 & 30.1601, 23.8829 & 22:40:00 & Rising & 18 \\
& 2006-09-25, 20:13:36 & 34.6437, 25.8576 & 22:32:09 & Rising & 19 \\
& 2006-09-26, 19:47:50 & 39.1949, 27.6340 & 22:24:36 & Rising & 20 \\
\hline
\multirow{3}{*}{
\begin{tabular}{l}
Suzaku/XIS+PIN\\(100023010)
\end{tabular}}
& 2005-09-15, 15:47:27 & 112.9348, 21.6094 & 23:19:10 & Rising & 21 \\
& 2005-09-15, 17:23:23 & 88.7894, 21.8078 & 23:18:32 & Rising & 21 \\
& 2005-09-15, 18:59:19 & 64.6462, 22.0065 & 23:17:53 & Rising & 21 \\
\hdashline
\multirow{1}{*}{
\begin{tabular}{l}
PIN only\\
\end{tabular}}
& 2005-09-15, 14:11:30 & 117.5 154.0 & 23:19:50 & Rising & 21\\
\hline
\multirow{2}{*}{
\begin{tabular}{l}
Suzaku/XIS+PIN\\(100023020)
\end{tabular}}
& 2005-09-15, 20:35:15 & 40.5040, 22.2049 & 23:17:15 & Rising & 21 \\
& 2005-09-16, 01:23:04 & 328.0701, 22.7902 & 23:15:20 & Rising & 21 \\
\hline
\multirow{11}{*}{
\begin{tabular}{l}
Suzaku/XIS+PIN\\(102019010)
\end{tabular}}
& 2007-03-20, 11:03:59 & 203.0962, 28.9984 & 00:36:21 & Setting & 22 \\
& 2007-03-20, 19:03:28 & 82.4316, 28.1415 &  00:33:11 & Setting & 22 \\
& 2007-03-20, 20:39:21 & 58.3004, 27.9608 &  00:32:32 & Setting & 22 \\
& 2007-03-20, 22:15:15 & 34.1662, 27.7856 &  00:31:54 & Setting & 22 \\
& 2007-03-20, 23:51:09 & 10.0323, 27.6087 &  00:31:15 & Setting & 22 \\
& 2007-03-21, 01:27:03 & 345.9020, 27.4311 & 00:30:38 & Setting & 22\\
& 2007-03-21, 03:02:56 & 321.7692, 27.2496 & 00:30:00 & Setting & 22 \\
& 2007-03-21, 04:38:50 & 297.6363, 27.0733 & 00:29:21 & Setting & 22 \\
& 2007-03-21, 06:14:44 & 273.5048, 26.8952 & 00:28:45 & Setting & 22 \\
& 2007-03-21, 07:50:38 & 249.3681, 26.7177 & 00:28:05 & Setting & 22 \\
& 2007-03-21, 09:26:31 & 225.2351, 26.5324 & 00:27:27 & Setting & 22 \\
\hdashline
\multirow{6}{*}{
\begin{tabular}{l}
PIN only\\
\end{tabular}}
& 2007-03-20, 12:39:53 & 178.9600, 28.8299 & 00:35:43 & Setting & 22\\
& 2007-03-20, 14:15:47 & 154.8306, 28.6595 & 00:35:05 & Setting & 22\\
& 2007-03-20, 15:51:40 & 130.6976, 28.4824 & 00:34:27 & Setting & 22\\
& 2007-03-20, 17:27:34 & 106.5661, 28.3117 & 00:33:48 & Setting & 22\\
& 2007-03-21, 11:02:25 & 201.1047, 26.3490 & 00:26:49 & Setting & 22\\
& 2007-03-21, 12:38:19 & 176.9683, 26.1683 & 00:26:11 & Setting & 22\\
\hline
\multirow{10}{*}{
\begin{tabular}{l}
Suzaku/XIS+PIN\\(103007010)
\end{tabular}}
& 2008-08-27, 08:53:45 & 240.7545, 8.8327 & 00:56:45 & Rising & 23 \\
& 2008-08-27, 10:29:38 & 216.6275, 9.0504 & 00:56:08 & Rising & 23 \\
& 2008-08-27, 12:05:31 & 192.5022, 9.2686 & 00:55:31 & Rising & 23 \\
& 2008-08-27, 13:41:25 & 168.3694, 9.4947 & 00:54:52 & Rising & 23 \\
& 2008-08-27, 15:17:18 & 144.2389, 9.7116 & 00:54:15 & Rising & 23 \\
& 2008-08-27, 16:53:11 & 120.1106, 9.9297 & 00:53:36 & Rising & 23 \\
& 2008-08-27, 18:29:05 & 95.9780, 10.1570 & 00:52:59 & Rising & 23 \\
& 2008-08-27, 20:04:58 & 71.8479, 10.3754 & 00:52:21 & Rising & 23 \\
& 2008-08-27, 21:40:51 & 47.7197, 10.5942 & 00:51:42 & Rising & 23 \\
& 2008-08-28, 04:04:24 & 311.2021, 11.4750& 00:49:11 & Rising & 23 \\
\hline
\multirow{11}{*}{
\begin{tabular}{l}
Suzaku/PIN (only) \\(104001010)
\end{tabular}}
& 2009-04-02, 02:28:52 & 312.5537, 15.8190 & 23:19:04 & Setting & 24 \\
& 2009-04-02, 05:40:37 & 264.3008, 15.3940 & 23:17:48 & Setting & 24\\
& 2009-04-02, 07:16:29 & 240.1773, 15.1828 & 23:17:11 & Setting & 24\\
& 2009-04-02, 08:52:22 & 216.0431, 14.9684 & 23:16:32 & Setting & 24\\
& 2009-04-02, 10:28:15 & 191.9123, 14.7529 & 23:15:52 & Setting & 24\\
& 2009-04-02, 12:04:08 & 167.7842, 14.5363 & 23:15:15 & Setting & 24\\
& 2009-04-02, 13:40:00 & 143.6593, 14.3257 & 23:14:38 & Setting & 24\\
& 2009-04-02, 15:15:53 & 119.5299, 14.1091 & 23:13:59 & Setting & 24\\
& 2009-04-02, 16:51:45 & 95.4068, 13.8977 &  23:13:22 & Setting & 24\\
& 2009-04-02, 18:27:38 & 71.2755, 13.6801 &  23:12:43 & Setting & 24\\
& 2009-04-02, 20:03:31 & 47.1460, 13.4604 &  23:12:06 & Setting & 24\\
& 2009-04-02, 21:39:24 & 23.0145, 13.2414 &  23:11:26 & Setting & 24\\
& 2009-04-02, 23:15:16 & 358.8897, 13.0297 & 23:10:49 & Setting & 24\\
& 2009-04-03, 00:51:09 & 334.7635, 12.8104 & 23:10:12 & Setting & 24\\
\hline
\multirow{14}{*}{
\begin{tabular}{l}
Suzaku/XIS+PIN\\(104001070)
\end{tabular}}
& 2010-02-23, 01:07:24 & 27.5234, 44.5019 &  02:57:29 & Setting & 25\\
& 2010-02-23, 02:43:15 & 3.4639, 44.4581 &   02:57:06 & Setting & 25\\
& 2010-02-23, 04:19:07 & 339.4214, 44.4512 & 02:56:48 & Setting & 25\\
& 2010-02-23, 05:54:58 & 315.3504, 44.4045 & 02:56:21 & Setting & 25\\
& 2010-02-23, 13:54:14 & 195.0371, 44.2031 & 02:54:21  & Setting & 25\\
& 2010-02-23, 15:30:05 & 170.9671, 44.1518 & 02:53:57 & Setting & 25\\
& 2010-02-23, 17:05:56 & 146.8931, 44.0990 & 02:53:30 & Setting & 25\\
& 2010-02-23, 18:41:48 & 122.8454, 44.0821 & 02:53:09 & Setting & 25\\
& 2010-02-23, 20:17:39 & 98.7736, 44.0294 &  02:52:44 & Setting & 25\\
& 2010-02-23, 21:53:30 & 74.6982, 43.9783 &  02:52:17 & Setting & 25\\
& 2010-02-23, 23:29:21 & 50.6253, 43.9263 &  02:51:51 & Setting & 25\\
& 2010-02-24, 01:05:12 & 26.5563, 43.8720 &  02:51:24 & Setting & 25\\
& 2010-02-24, 02:41:03 & 2.4805, 43.8172 &   02:50:57 & Setting & 25\\
& 2010-02-24, 04:16:54 & 338.4038, 43.7578 & 02:50:30 & Setting & 25\\
\hline
\multirow{7}{*}{
\begin{tabular}{l}
Suzaku/XIS+PIN\\(105002010)
\end{tabular}}
& 2010-04-06, 00:13:27 & 132.5794, 44.7686 & 09:03:46 & Rising & 26\\
& 2010-04-06, 01:49:18 & 108.5461, 44.7772 & 09:03:29 & Rising & 26\\
& 2010-04-06, 03:25:10 & 84.5450, 44.7403 & 09:03:20  & Rising & 26\\
& 2010-04-06, 05:01:02 & 60.5457, 44.6994 & 09:03:11  & Rising & 26\\
& 2010-04-06, 06:36:54 & 36.5486, 44.6582 & 09:03:05  & Rising & 26\\
& 2010-04-06, 08:12:45 & 12.5208, 44.6633 & 09:02:48  & Rising & 26\\
& 2010-04-06, 09:48:37 & 348.5188, 44.6257 & 09:02:40 & Rising & 26 \\
\hdashline
\multirow{5}{*}{
\begin{tabular}{l}
PIN only\\
\end{tabular}}
& 2010-04-05, 16:14:09 & 252.6709, 44.8427 & 09:04:50 & Rising & 26\\
& 2010-04-05, 17:50:00 & 228.6303, 44.8519 & 09:04:31 & Rising & 26\\
& 2010-04-05, 19:25:52 & 204.6280, 44.8210 & 09:04:22 & Rising & 26\\
& 2010-04-05, 21:01:43 & 180.5906, 44.8324 & 09:04:04 & Rising & 26\\
& 2010-04-05, 22:37:35 & 156.5839, 44.8016 & 09:03:55 & Rising & 26\\
\hline
\multirow{18}{*}{
\begin{tabular}{l}
Suzaku/XIS+PIN\\(105029010)
\end{tabular}}
& 2011-03-21, 19:41:41 & 78.7092, 39.2653 &  00:56:30 & Setting & 27\\
& 2011-03-21, 21:17:33 & 54.7942, 39.4346 &  00:56:43 & Setting & 27\\
& 2011-03-21, 22:53:25 & 30.8867, 39.6003 &  00:56:57  & Setting & 27\\
& 2011-03-22, 00:29:18 & 7.0040, 39.8187 &   00:57:17 & Setting & 27\\
& 2011-03-22, 02:05:10 & 343.0985, 39.9859 & 00:57:33 & Setting & 27\\
& 2011-03-22, 03:41:02 & 319.1887, 40.1518 & 00:57:47 & Setting & 27\\
& 2011-03-22, 05:16:54 & 295.2737, 40.3134 & 00:57:59 & Setting & 27\\
& 2011-03-22, 10:04:30 & 223.5321, 40.7892 & 00:58:36 & Setting & 27\\
& 2011-03-22, 11:40:22 & 199.6170, 40.9492 & 00:58:50 & Setting & 27\\
& 2011-03-22, 13:16:13 & 175.6760, 41.0566 & 00:58:54 & Setting & 27\\
& 2011-03-22, 14:52:05 & 151.7591, 41.2150 & 00:59:07 & Setting & 27\\
& 2011-03-22, 16:27:57 & 127.8495, 41.3677 & 00:59:20 & Setting & 27\\
& 2011-03-22, 18:03:48 & 103.9000, 41.4732 & 00:59:23 & Setting & 27\\
& 2011-03-22, 19:39:40 & 79.9866, 41.6262 &  00:59:35 & Setting & 27\\
& 2011-03-22, 21:15:32 & 56.0717, 41.7799 &  00:59:48 & Setting & 27\\
& 2011-03-22, 22:51:23 & 32.1257, 41.8849 &  00:59:53 & Setting & 27\\
& 2011-03-23, 00:27:15 & 8.2138, 42.0381 &   01:00:06 & Setting & 27\\
& 2011-03-23, 02:03:06 & 344.2684, 42.1442 & 01:00:10 & Setting & 27\\
\hline
\multirow{3}{*}{
\begin{tabular}{l}
Suzaku/XIS+PIN\\(106012010)
\end{tabular}}
& 2011-09-01, 06:39:28 & 252.0472, 42.5591 & 23:27:38 & Rising & 28\\
& 2011-09-01, 07:45:03 & 79.6606, 5.5418 &   13:03:41 & Setting & 29\\
& 2011-09-01, 08:15:17 & 228.0969, 42.4752 & 23:27:39 & Rising & 28 \\
& 2011-09-01, 09:51:07 & 204.1749, 42.3442 & 23:27:47 & Rising & 28\\
& 2011-09-01, 10:56:49 & 31.8986, 6.3004 &   13:04:23 & Setting & 29\\
& 2011-09-01, 11:26:56 & 180.2215, 42.2622 & 23:27:48 & Rising & 28\\
& 2011-09-01, 13:02:46 & 156.3045, 42.1316 & 23:27:59 & Rising & 28\\
& 2011-09-01, 14:38:36 & 132.3875, 41.9985 & 23:28:08 & Rising & 28\\
& 2011-09-01, 16:14:26 & 108.4658, 41.8605 & 23:28:17 & Rising & 28\\
& 2011-09-01, 17:50:15 & 84.5215, 41.7703 &  23:28:19 & Rising & 28\\
& 2011-09-01, 19:26:05 & 60.6072, 41.6296 &  23:28:30 & Rising & 28\\
& 2011-09-01, 20:32:07 & 248.6250, 8.5945 &  13:06:37 & Setting & 29\\
& 2011-09-01, 21:01:55 & 36.6872, 41.4873 &  23:28:38 & Rising & 28\\
& 2011-09-01, 22:08:00 & 224.7420, 8.9729 &  13:06:57 & Setting & 29\\
& 2011-09-01, 22:37:45 & 12.7723, 41.3508 &  23:28:50 & Rising & 28\\
& 2011-09-01, 23:43:53 & 200.8643, 9.3497 &  13:07:19 & Setting & 29\\
& 2011-09-02, 00:13:35 & 348.8518, 41.2116 & 23:28:59 & Rising & 28\\
& 2011-09-02, 01:19:46 & 176.9834, 9.7286 &  13:07:41 & Setting & 29\\
& 2011-09-02, 02:55:40 & 153.1248, 10.1580 & 13:08:08 & Setting & 29\\
& 2011-09-02, 04:31:33 & 129.2479, 10.5358 & 13:08:31 & Setting & 29\\
& 2011-09-02, 06:07:26 & 105.3699, 10.9140 & 13:08:53 & Setting & 29\\
\hline
\multirow{14}{*}{
\begin{tabular}{l}
Suzaku/XIS+PIN\\(106014010)
\end{tabular}}
& 2012-03-14, 01:27:52 & 348.9473, 21.5192 & 00:43:39  & Setting &  30\\
& 2012-03-14, 03:03:35 & 324.8601, 21.3204 & 00:43:00 & Setting & 30 \\
& 2012-03-14, 04:39:18 & 300.7737, 21.1207 & 00:42:23 & Setting & 30 \\
& 2012-03-14, 06:15:01 & 276.6836, 20.9213 & 00:41:44 & Setting & 30\\
& 2012-03-14, 07:50:44 & 252.5936, 20.7208 & 00:41:06 & Setting & 30\\
& 2012-03-14, 09:26:27 & 228.5048, 20.5185 & 00:40:27 & Setting & 30\\
& 2012-03-14, 11:02:10 & 204.4156, 20.3156 & 00:39:48 & Setting & 30\\
& 2012-03-14, 12:37:53 & 180.3265, 20.1133 & 00:39:11 & Setting & 30\\
& 2012-03-14, 14:13:36 & 156.2397, 19.9099 & 00:38:32 & Setting & 30\\
& 2012-03-14, 15:49:19 & 132.1504, 19.7072 & 00:37:54 & Setting & 30\\
& 2012-03-14, 17:25:02 & 108.0625, 19.5037 & 00:37:15 & Setting & 30\\
& 2012-03-14, 19:00:45 & 83.9744, 19.2992 &  00:36:38  & Setting & 30\\
& 2012-03-14, 20:36:28 & 59.8841, 19.0937 &  00:36:00 & Setting & 30\\
& 2012-03-14, 22:12:11 & 35.7953, 18.8861 &  00:35:21  & Setting & 30\\
\hline
\multirow{12}{*}{
\begin{tabular}{l}
Suzaku/XIS+PIN\\(107011010)
\end{tabular}}
& 2012-09-26, 06:38:44 & 225.9450, 43.6655 & 21:42:29 & Rising & 31\\
& 2012-09-26, 08:14:24 & 201.9427, 43.6907 & 21:42:10 & Rising & 31\\
& 2012-09-26, 09:50:03 & 177.9161, 43.7499 & 21:41:42 & Rising & 31\\
& 2012-09-26, 11:25:42 & 153.8869, 43.8101 & 21:41:13 & Rising & 31\\
& 2012-09-26, 13:01:22 & 129.8874, 43.8355 & 21:40:54 & Rising & 31\\
& 2012-09-26, 14:37:01 & 105.8578, 43.8916 & 21:40:26 & Rising & 31\\
& 2012-09-26, 16:12:40 & 81.8353, 43.9488 &  21:40:00 & Rising & 31\\
& 2012-09-26, 17:48:19 & 57.8115, 44.0017 &  21:39:33 & Rising & 31 \\
& 2012-09-26, 19:23:59 & 33.8157, 44.0155 &  21:39:14 & Rising & 31\\
& 2012-09-26, 20:59:38 & 9.7952, 44.0657 &   21:38:48 & Rising & 31\\
& 2012-09-26, 22:35:17 & 345.7727, 44.1193 & 21:38:21 & Rising & 31\\
& 2012-09-27, 04:57:55 & 249.7428, 44.2478 & 21:36:52 & Rising & 31\\
\hline
\multirow{6}{*}{
\begin{tabular}{l}
Suzaku/XIS+PIN\\(107012010)
\end{tabular}}
& 2013-02-27, 01:09:14 & 176.7227, 6.7619 & 12:56:07 & Rising & 32\\
& 2013-02-27, 02:44:51 & 152.6637, 6.9839 & 12:55:30 & Rising & 32\\
& 2013-02-27, 04:20:28 & 128.6018, 7.2055 & 12:54:51 & Rising & 32\\
& 2013-02-27, 05:56:05 & 104.5407, 7.4277 & 12:54:14 & Rising & 32\\
& 2013-02-27, 07:31:42 & 80.4816, 7.6515 &  12:53:36 & Rising & 32\\
& 2013-02-27, 09:07:19 & 56.4198, 7.8743 &  12:52:59 & Rising & 32\\
& 2013-02-27, 17:05:23 & 296.1170, 8.9776 & 12:49:50 & Rising & 32\\
& 2013-02-27, 18:41:00 & 272.0577, 9.2017 & 12:49:13 & Rising & 32\\
& 2013-02-27, 20:16:37 & 247.9946, 9.4233 & 12:48:34 & Rising & 32\\
\hdashline
\multirow{1}{*}{
\begin{tabular}{l}
PIN only \\
\end{tabular}}
& 2013-02-27, 21:52:14 & 223.9340, 9.6449 & 12:47:57 & Rising & 32\\
\hline
\multirow{14}{*}{
\begin{tabular}{l}
Suzaku/XIS+PIN\\(108011010)
\end{tabular}}
& 2013-09-30, 11:41:53 & 357.3104, 19.7266 & 11:31:06 & Setting & 33\\
& 2013-09-30, 13:17:25 & 333.2719, 19.5224 & 11:30:29 & Setting & 33\\
& 2013-09-30, 14:52:57 & 309.2295, 19.3195 & 11:29:52  & Setting & 33\\
& 2013-09-30, 16:28:29 & 285.1842, 19.1169 & 11:29:12 & Setting & 33\\
& 2013-09-30, 18:04:01 & 261.1428, 18.9115 & 11:28:35 & Setting & 33\\
& 2013-09-30, 19:39:34 & 237.0956, 18.7033 & 11:27:56 & Setting & 33\\
& 2013-09-30, 21:15:06 & 213.0517, 18.4971 & 11:27:18  & Setting & 33\\
& 2013-09-30, 22:50:38 & 189.0108, 18.2901 & 11:26:39 & Setting & 33\\
& 2013-10-01, 00:26:10 & 164.9684, 18.0839 & 11:26:02 & Setting & 33\\
& 2013-10-01, 02:01:42 & 140.9276, 17.8767 & 11:25:23 & Setting & 33\\
& 2013-10-01, 03:37:15 & 116.8792, 17.6673 & 11:24:46 & Setting & 33\\
& 2013-10-01, 05:12:47 & 92.8361, 17.4602 &  11:24:06 & Setting & 33\\
& 2013-10-01, 06:48:19 & 68.7952, 17.2506 &  11:23:29 & Setting & 33\\
& 2013-10-01, 08:23:51 & 44.7519, 17.0404 &  11:22:50 & Setting & 33\\
\hdashline
\multirow{1}{*}{
\begin{tabular}{l}
PIN only \\
\end{tabular}}
& 2013-10-01, 09:59:23 & 20.7104, 16.8293 & 11:22:13 & Setting & 33\\
\hline
\multirow{11}{*}{
\begin{tabular}{l}
Suzaku/XIS+PIN\\(108012010)
\end{tabular}}
& 2014-03-06, 01:51:23 & 5.2762, 44.0279 &   02:12:29 & Setting & 34\\
& 2014-03-06, 03:26:48 & 341.3101, 43.9703 & 02:12:02 & Setting & 34\\
& 2014-03-06, 05:02:14 & 317.3638, 43.9483 & 02:11:40 & Setting & 34\\
& 2014-03-06, 12:59:21 & 197.5711, 43.7266 & 02:09:38 & Setting & 34\\
& 2014-03-06, 14:34:46 & 173.6008, 43.6648 & 02:09:10 & Setting & 34\\
& 2014-03-06, 16:10:11 & 149.6319, 43.5993 & 02:08:42 & Setting & 34\\
& 2014-03-06, 17:45:37 & 125.6839, 43.5707 & 02:08:21 & Setting & 34\\
& 2014-03-06, 19:21:02 & 101.7124, 43.5079 & 02:07:51 & Setting & 34\\
& 2014-03-06, 20:56:28 & 77.7640, 43.4797 &  02:07:31 & Setting & 34\\
& 2014-03-06, 22:31:53 & 53.7919, 43.4175 &  02:07:03 & Setting & 34\\
& 2014-03-07, 00:07:18 & 29.8190, 43.3556 &  02:06:33 & Setting & 34\\
\hline
\multirow{4}{*}{
\begin{tabular}{l}
NuSTAR/FPM\\(10013021002)
\end{tabular}}
& 2012-09-20, 03:14:44 & 299.2689, 15.5307 & 23:11:48 & Rising & 35 \\
& 2012-09-20, 04:16:38 & 112.4618, 5.5680 &  11:46:27 & Setting & 36 \\
& 2012-09-20, 04:51:46 & 274.9422, 15.5296 & 23:11:31 & Rising & 35 \\
& 2012-09-20, 05:53:41 & 88.1595, 5.6261 &  11:46:18  & Setting & 36 \\
\hline
\multirow{4}{*}{
\begin{tabular}{l}
NuSTAR/FPM\\(10013024002)
\end{tabular}}
& 2012-09-20, 11:19:56 & 177.6637, 15.4670 & 23:10:35 & Rising & 37 \\
& 2012-09-20, 12:21:52 & 350.9328, 5.8355 & 11:45:34 & Setting & 38 \\
& 2012-09-20, 12:56:58 & 153.3413, 15.4654 & 23:10:18  & Rising & 37 \\
& 2012-09-20, 13:58:54 & 326.6212, 5.8660 & 11:45:23 & Setting & 38 \\
\hline
\multirow{4}{*}{
\begin{tabular}{l}
NuSTAR/FPM\\(10013022002)
\end{tabular}}
& 2012-09-20, 19:25:08 & 55.7379, 15.3000 & 23:08:04 & Rising & 39\\
& 2012-09-20, 20:27:05 & 229.0669, 6.1742 & 11:43:21 & Setting & 40\\
& 2012-09-20, 21:02:10 & 31.4155, 15.2962 & 23:07:49 & Rising & 39\\
& 2012-09-20, 22:04:08 & 204.7604, 6.2364 & 11:43:09 & Setting & 40\\
\hline
\multirow{4}{*}{
\begin{tabular}{l}
NuSTAR/FPM\\(10013022004)
\end{tabular}}
& 2012-09-21, 03:30:20 & 294.4665, 15.3093 & 23:08:11 & Rising & 41 \\
& 2012-09-21, 04:32:19 & 107.8726, 6.3589 &  11:43:48 & Setting & 42 \\
& 2012-09-21, 05:07:22 & 270.1443, 15.3033 & 23:07:55 & Rising & 41 \\
& 2012-09-21, 06:09:22 & 83.5663, 6.4222 &  11:43:36  & Setting & 42 \\
\hline
\multirow{2}{*}{
\begin{tabular}{l}
NuSTAR/FPM\\(10013022006)
\end{tabular}}
& 2012-09-21, 16:26:39 & 99.9226, 15.1632 & 23:06:20 & Rising & 43 \\
& 2012-09-21, 18:03:42 & 75.6101, 15.1242 & 23:06:07 & Rising & 43 \\
\hline
& 2013-03-09, 07:46:38 & 68.9599, 2.5936 & 12:22:27 & Rising & 44 \\
\multirow{5}{*}{
\begin{tabular}{l}
NuSTAR/FPM\\(80001022002)
\end{tabular}}
& 2013-03-09, 08:48:05 & 240.6845, 14.6767 & 00:50:48 & Setting & 45 \\
& 2013-03-09, 09:23:38 & 44.6454, 2.5835 &  12:22:11  & Rising & 44 \\
& 2013-03-09, 10:25:05 & 216.3539, 14.6495 & 00:50:29& Setting & 45 \\
& 2013-03-09, 11:00:39 & 20.3403, 2.5505 &  12:22:00  & Rising & 44 \\
& 2013-03-09, 12:02:05 & 192.0277, 14.6201 & 00:50:11& Setting & 45 \\
\hline
\multirow{4}{*}{
\begin{tabular}{l}
NuSTAR/FPM\\(10013037002)
\end{tabular}}
& 2013-04-03, 09:09:49 & 18.2802, 15.0293 & 10:22:55  & Rising & 46 \\
& 2013-04-03, 10:11:17 & 190.2760, 2.8922 &  22:52:23 & Setting & 47 \\
& 2013-04-03, 10:46:49 & 353.9586, 15.0456 & 10:22:38 & Rising & 46 \\
& 2013-04-03, 11:48:17 & 165.9694, 2.9140 &  22:52:09 & Setting & 47 \\
\hline
\multirow{4}{*}{
\begin{tabular}{l}
NuSTAR/FPM\\(10013037004)
\end{tabular}}
& 2013-04-04, 11:01:45 & 349.1032, 15.3330 & 10:18:09 & Rising & 48 \\
& 2013-04-04, 12:03:17 & 161.3774, 3.2953 &  22:48:47 & Setting & 49 \\
& 2013-04-04, 12:38:45 & 324.7830, 15.3422 & 10:17:52 & Rising & 48 \\
& 2013-04-04, 13:40:17 & 137.0754, 3.3221 &  22:48:34 & Setting & 49 \\
\hline
\multirow{4}{*}{
\begin{tabular}{l}
NuSTAR/FPM\\(10013037006)
\end{tabular}}
& 2013-04-05, 06:25:42 & 57.2489, 15.4965 & 10:14:40 & Rising & 50 \\
& 2013-04-05, 07:27:18 & 229.7341, 3.6801 & 22:46:14 & Setting & 51 \\
& 2013-04-05, 08:02:42 & 32.9296, 15.5003 & 10:14:25 & Rising & 50 \\
& 2013-04-05, 09:04:18 & 205.4322, 3.7116 & 22:46:00 & Setting & 51 \\
\hline
\multirow{4}{*}{
\begin{tabular}{l}
NuSTAR/FPM\\(10013038002)
\end{tabular}}
& 2013-04-08, 12:01:33 & 330.1119, 15.4659 & 10:01:59 & Rising & 52\\
& 2013-04-08, 13:03:25 & 143.3684, 5.7677 & 22:36:52  & Setting & 53\\
& 2013-04-08, 13:38:33 & 305.8007, 15.4487 & 10:01:45 & Rising & 52\\
& 2013-04-08, 14:40:25 & 119.0723, 5.8124 & 22:36:42  & Setting & 53\\
\hline
\multirow{4}{*}{
\begin{tabular}{l}
NuSTAR/FPM\\(10013038004)
\end{tabular}}
& 2013-04-09, 21:58:31 & 179.6391, 15.0960 & 09:57:03 & Rising & 54\\
& 2013-04-09, 23:00:29 & 353.1661, 6.8803 &  22:33:07 & Setting & 55\\
& 2013-04-09, 23:35:31 & 155.3323, 15.0710 & 09:56:49 & Rising & 54\\
& 2013-04-10, 00:37:29 & 328.8659, 6.9306 &  22:32:56 & Setting & 55\\
\hline
\multirow{4}{*}{
\begin{tabular}{l}
NuSTAR/FPM\\(10013037008)
\end{tabular}}
& 2013-04-18, 09:40:41 & 357.8577, 9.2175 &  09:32:06  & Rising & 56\\
& 2013-04-18, 10:42:49 & 171.7134, 13.7309 & 22:09:39 & Setting & 57\\ 
& 2013-04-18, 11:17:41 & 333.5635, 9.1638 &  09:31:55 & Rising & 56\\
& 2013-04-18, 12:19:49 & 147.4151, 13.7684 & 22:09:28 & Setting & 57\\
\hline
\multirow{4}{*}{
\begin{tabular}{l}
NuSTAR/FPM\\(10002001002)
\end{tabular}}
& 2013-09-02, 04:29:03 & 298.6616, 15.4632 & 00:23:40  & Rising & 58 \\
& 2013-09-02, 05:30:37 & 111.1456, 3.6867 &  12:55:10 & Setting & 59 \\
& 2013-09-02, 06:06:01 & 274.3493, 15.4756 & 00:23:24 & Rising & 58 \\
& 2013-09-02, 07:07:35 & 86.8473, 3.7114 &   12:54:57 & Setting & 59 \\
\hline
\multirow{4}{*}{
\begin{tabular}{l}
NuSTAR/FPM\\(10002001004)
\end{tabular}}
& 2013-09-03, 06:20:32 & 269.6793, 15.5823 & 00:19:14 & Rising & 60 \\
& 2013-09-03, 07:22:11 & 82.4347, 4.2782 &   12:51:55 & Setting & 61 \\
& 2013-09-03, 07:57:30 & 245.3676, 15.5876 & 00:18:57 & Rising & 60 \\
& 2013-09-03, 08:59:09 & 58.1361, 4.3083 &   12:51:41 & Setting & 61 \\
\hline
\multirow{8}{*}{
\begin{tabular}{l}
NuSTAR/FPM\\(10002001008)
\end{tabular}}
& 2014-10-02, 03:23:18 & 286.5470, 13.8173 & 22:29:29 & Rising & 62 \\
& 2014-10-02, 04:24:24 & 97.8869, 2.3433 &   10:55:55 & Setting & 63 \\
& 2014-10-02, 05:00:08 & 262.2535, 13.8614 & 22:29:08 & Rising & 62 \\
& 2014-10-02, 06:01:15 & 73.6147, 2.3598 &   10:55:41 & Setting & 63 \\
& 2014-10-02, 06:36:58 & 237.9601, 13.9053 & 22:28:48 & Rising & 62 \\
& 2014-10-02, 07:38:05 & 49.3375, 2.3531 &   10:55:26 & Setting & 63 \\
& 2014-10-02, 08:13:49 & 213.6728, 13.9254 & 22:28:30 & Rising & 62 \\
& 2014-10-02, 09:14:56 & 25.0655, 2.3706 &   10:55:10 & Setting & 63 \\
\hline
\multirow{5}{*}{
\begin{tabular}{l}
NuSTAR/FPM\\(10002001009)
\end{tabular}}
& 2016-04-01, 07:20:28 & 47.9118, 13.9393 & 10:32:06 & Rising & 64 \\
& 2016-04-01, 08:21:27 & 219.3273, 2.3173 & 22:58:45 & Setting & 65 \\
& 2016-04-01, 08:57:09 & 23.6581, 13.9778 & 10:31:45 & Rising & 64 \\
& 2016-04-01, 09:58:08 & 195.0895, 2.3163 & 22:58:28 & Setting & 65 \\
& 2016-04-01, 10:33:50 & 359.4044, 14.0159 & 10:31:27 & Rising & 64 \\
& 2016-04-01, 11:34:50 & 170.8569, 2.3396 & 22:58:14 & Setting & 65 \\
& 2016-04-01, 12:10:32 & 335.1568, 14.0299 & 10:31:09 & Rising & 64 \\
& 2016-04-01, 13:11:31 & 146.6191, 2.3395 & 22:57:59 & Setting & 65 \\
\hline
\multirow{4}{*}{
\begin{tabular}{l}
NuSTAR/FPM\\(10202001006)
\end{tabular}}
& 2017-02-19, 19:49:00 & 92.6967, 14.5072 & 01:59:47 & Setting & 66 \\
& 2017-02-19, 21:25:40 & 68.4563, 14.4823 & 01:59:29 & Setting & 66 \\
& 2017-02-19, 23:02:20 & 44.2117, 14.4588 & 01:59:09 & Setting & 66 \\
& 2017-02-20, 00:39:00 & 19.9671, 14.4348 & 01:58:51 & Setting & 66 \\
\hline
\multirow{8}{*}{
\begin{tabular}{l}
NuSTAR/FPM\\(10302001002)
\end{tabular}}
& 2017-09-02, 07:37:03 & 256.2480, 4.2871 & 00:42:02  & Rising & 67 \\
& 2017-09-02, 08:38:20 & 69.0233, 15.4090 & 13:14:25 & Setting & 68 \\
& 2017-09-02, 09:13:43 & 232.0362, 4.2403 & 00:41:51 & Rising & 67 \\
& 2017-09-02, 10:15:00 & 44.7982, 15.4209 & 13:14:11 & Setting & 68 \\
& 2017-09-02, 10:50:23 & 207.8203, 4.1923 & 00:41:39 & Rising & 67 \\
& 2017-09-02, 11:51:39 & 20.5630, 15.4056 & 13:13:54 & Setting & 68 \\
& 2017-09-02, 12:27:03 & 183.6087, 4.1462 & 00:41:29 & Rising & 67 \\
& 2017-09-02, 13:28:19 & 356.3337, 15.4183 & 13:13:38 & Setting & 68 \\
\hline
\multirow{12}{*}{
\begin{tabular}{l}
NuSTAR/FPM\\(10302001004)
\end{tabular}}
& 2017-09-29, 18:29:01 & 62.0202, 14.2942 & 22:37:05 & Rising & 69 \\
& 2017-09-29, 19:30:02 & 233.6046, 2.3753 & 11:04:27 & Setting & 70  \\
& 2017-09-29, 20:05:40 & 37.7792, 14.3291 & 22:36:47 & Rising & 69 \\
& 2017-09-29, 21:06:41 & 209.3772, 2.3787 & 11:04:10 & Setting & 70 \\
& 2017-09-29, 21:42:20 & 13.5400, 14.3374 & 22:36:29 & Rising & 69 \\
& 2017-09-29, 22:43:21 & 185.1551, 2.4071 & 11:03:57 & Setting & 70 \\
& 2017-09-29, 23:18:59 & 349.2991, 14.3715 & 22:36:10 & Rising & 69 \\
& 2017-09-30, 00:20:00 & 160.9278, 2.4113 & 11:03:42  & Setting & 70 \\
& 2017-09-30, 00:55:38 & 325.0540, 14.4036 & 22:35:49 & Rising & 69 \\
& 2017-09-30, 01:56:40 & 136.7058, 2.4407 & 11:03:29 & Setting & 70 \\
& 2017-09-30, 02:32:17 & 300.8132, 14.4371 & 22:35:31 & Rising & 69 \\
& 2017-09-30, 03:33:19 & 112.4785, 2.4457 & 11:03:12 & Setting & 70 \\
\hline
\multirow{6}{*}{
\begin{tabular}{l}
NuSTAR/FPM\\(10402001004)
\end{tabular}}
& 2018-03-13, 00:18:24 & 174.8090, 9.5597 &  11:57:38 & Rising & 71 \\
& 2018-03-13, 01:20:06 & 348.7072, 13.2345 & 00:34:54 & Setting & 72 \\
& 2018-03-13, 01:55:04 & 150.6043, 9.4813 & 11:57:29  & Rising & 71 \\
& 2018-03-13, 02:56:45 & 324.4931, 13.2702 & 00:34:42 & Setting & 72 \\
& 2018-03-13, 03:31:43 & 126.3984, 9.4373 &  11:57:18 & Rising & 71 \\
& 2018-03-13, 04:33:25 & 300.2886, 13.3366 & 00:34:33 & Setting & 72 \\
\hline
\multirow{10}{*}{
\begin{tabular}{l}
NuSTAR/FPM\\(10402001008)
\end{tabular}}
& 2018-03-14, 11:44:51 & 2.2749, 8.2658 &  11:53:56   & Rising & 73\\
& 2018-03-14, 12:46:29 & 176.0298, 14.1074 & 00:30:36 & Setting & 74\\
& 2018-03-14, 13:21:30 & 338.0655, 8.2195 &  11:53:44 & Rising & 73\\
& 2018-03-14, 14:23:08 & 151.8128, 14.1372 & 00:30:23 & Setting & 74 \\
& 2018-03-14, 14:58:10 & 313.8656, 8.1428 &  11:53:36 & Rising & 73\\
& 2018-03-14, 15:59:47 & 127.5960, 14.1665 & 00:30:10 & Setting & 74\\
& 2018-03-14, 16:34:49 & 289.6562, 8.0966 & 11:53:26  & Rising & 73\\
& 2018-03-14, 17:36:27 & 103.3889, 14.2259 & 00:30:00 & Setting & 74\\
& 2018-03-14, 18:11:29 & 265.4564, 8.0200 & 11:53:17  & Rising & 73\\
& 2018-03-14, 19:13:06 & 79.1722, 14.2544 & 00:29:47 & Setting & 74\\
\hline
\multirow{6}{*}{
\begin{tabular}{l}
NuSTAR/FPM\\(10502001006)
\end{tabular}}
& 2019-03-11, 05:34:03 & 100.7470, 2.6298 & 12:17:2  & Rising & 75 \\
& 2019-03-11, 06:34:41 & 271.1959, 11.1829 & 00:39:28 & Setting & 76 \\
& 2019-03-11, 07:10:41 & 76.5135, 2.6532 &  12:16:44  & Rising & 75 \\
& 2019-03-11, 08:11:19 & 246.9517, 11.1325 & 00:39:06 & Setting & 76 \\
& 2019-03-11, 08:47:19 & 52.2800, 2.6770 &   12:16:26 & Rising & 75 \\
& 2019-03-11, 09:47:57 & 222.7075, 11.0821 & 00:38:45 & Setting & 76 \\
\hline
\multirow{4}{*}{
\begin{tabular}{l}
NuSTAR/FPM\\(10502001008)
\end{tabular}}
& 2019-03-11, 18:27:08 & 266.5429, 2.7104 & 12:13:18 & Rising & 77\\
& 2019-03-11, 19:27:45 & 76.8998, 10.8750 & 00:35:20 & Setting & 78 \\
& 2019-03-11, 20:03:46 & 242.3092, 2.7370 & 12:13:00 & Rising & 77\\
& 2019-03-11, 21:04:23 & 52.6511, 10.8252 & 00:34:59 & Setting & 78 \\
\hline
\multirow{14}{*}{
\begin{tabular}{l}
NuSTAR/FPM\\(10502001013)
\end{tabular}}
& 2019-08-29, 10:10:21 & 217.6229, 15.3713 & 00:40:50 & Rising & 79 \\
& 2019-08-29, 11:11:40 & 30.3802, 4.3013 &  13:13:10  & Setting & 80 \\
& 2019-08-29, 11:46:59 & 193.3996, 15.3712 & 00:40:33 & Rising & 79 \\
& 2019-08-29, 12:48:18 & 6.1695, 4.3354 &   13:12:57  & Setting &  80\\
& 2019-08-29, 13:23:37 & 169.1763, 15.3705 & 00:40:18 & Rising & 79 \\
& 2019-08-29, 14:24:56 & 341.9589, 4.3697 & 13:12:46  & Setting & 80 \\
& 2019-08-29, 15:00:15 & 144.9487, 15.3677 & 00:40:01 & Rising & 79 \\
& 2019-08-29, 16:01:35 & 317.7577, 4.4317 & 13:12:36  & Setting & 80 \\
& 2019-08-29, 16:36:53 & 120.7253, 15.3661 & 00:39:47 & Rising & 79 \\
& 2019-08-29, 17:38:13 & 293.5470, 4.4667 &  13:12:24 & Setting & 80 \\
& 2019-08-29, 18:13:31 & 96.5019, 15.3640 & 00:39:30 & Rising & 79 \\
& 2019-08-29, 19:14:51 & 269.3362, 4.5021 &  13:12:11 & Setting & 80 \\
& 2019-08-29, 19:50:09 & 72.2785, 15.3615 & 00:39:15  & Rising & 79 \\
& 2019-08-29, 20:51:30 & 245.1349, 4.5651 & 13:12:02  & Setting & 80 \\
\hline
\multirow{14}{*}{
\begin{tabular}{l}
NuSTAR/FPM\\(10502001015)
\end{tabular}}
& 2019-08-30, 08:43:13 & 238.4968, 15.3279 & 00:37:12& Rising & 81 \\
& 2019-08-30, 09:44:37 & 51.4801, 4.9439 &  13:10:32  & Setting & 82 \\
& 2019-08-30, 10:19:51 & 214.2773, 15.3231 & 00:36:56& Rising & 81 \\
& 2019-08-30, 11:21:15 & 27.2690, 4.9821 &   13:10:18 & Setting & 82 \\
& 2019-08-30, 11:56:30 & 190.0638, 15.2882 & 00:36:45& Rising & 81 \\
& 2019-08-30, 12:57:53 & 3.0620, 5.0188 &   13:10:06  & Setting & 82 \\
& 2019-08-30, 13:33:08 & 165.8400, 15.2806 & 00:36:29& Rising & 81 \\
& 2019-08-30, 14:34:32 & 338.8603, 5.0855 &  13:09:57 & Setting & 82 \\
& 2019-08-30, 15:09:46 & 141.6205, 15.2742 & 00:36:14& Rising & 81 \\
& 2019-08-30, 16:11:10 & 314.6490, 5.1244 &  13:09:44 & Setting & 82 \\
& 2019-08-30, 16:46:24 & 117.4009, 15.2674 & 00:36:0& Rising & 81 \\
& 2019-08-30, 17:47:48 & 290.4419, 5.1619 &  13:9:33 & Setting & 82 \\
& 2019-08-30, 18:23:02 & 93.1771, 15.2584 & 00:35:44 & Rising &  81\\
& 2019-08-30, 19:24:27 & 266.2401, 5.2296 &  13:09:24 & Setting & 82 \\
\hline
\multirow{6}{*}{
\begin{tabular}{l}
NuSTAR/FPM\\(10602002002)
\end{tabular}}
& 2020-02-27, 01:52:08 & 166.8508, 5.6676 & 12:59:31 & Rising & 83\\
& 2020-02-27, 02:52:38 & 336.6918, 6.6537 & 01:19:23 & Setting & 84\\
& 2020-02-27, 03:28:45 & 142.6076, 5.7239 & 12:59:09 & Rising & 83\\
& 2020-02-27, 04:29:15 & 312.4440, 6.5964 & 01:19:00& Setting & 84\\
& 2020-02-27, 05:05:23 & 118.3693, 5.7626 & 12:58:50 & Rising & 83\\
& 2020-02-27, 06:05:53 & 288.2054, 6.5555 & 01:18:42 & Setting & 84\\
\hline
\multirow{6}{*}{
\begin{tabular}{l}
NuSTAR/FPM\\(10602002004)
\end{tabular}}
& 2020-02-28, 00:24:52 & 187.4582, 6.3583 & 12:54:41 & Rising & 85\\
& 2020-02-28, 01:25:22 & 357.2962, 5.9578 & 01:14:32 & Setting & 86\\
& 2020-02-28, 02:01:29 & 163.2149, 6.4171 & 12:54:20 & Rising & 85\\
& 2020-02-28, 03:01:59 & 333.0528, 5.9013 & 01:14:11 & Setting & 86\\
& 2020-02-28, 03:38:07 & 138.9766, 6.4582 & 12:54:01 & Rising & 85\\
& 2020-02-28, 04:38:37 & 308.8104, 5.8645 & 01:13:50  & Setting & 86\\
\hline
\multirow{6}{*}{
\begin{tabular}{l}
NuSTAR/FPM\\(10602002006)
\end{tabular}}
& 2020-08-28, 09:35:02 & 231.5389, 3.8916 & 01:01:10 & Rising & 87 \\
& 2020-08-28, 10:36:17 & 44.1504, 15.3800 & 13:32:53 & Setting & 88 \\
& 2020-08-28, 11:11:39 & 207.3275, 3.8718 & 01:00:56& Rising & 87 \\
& 2020-08-28, 12:12:54 & 19.9219, 15.3664 & 13:32:35 & Setting & 88 \\
& 2020-08-28, 12:48:17 & 183.1217, 3.8242 & 01:00:46 & Rising & 87 \\
& 2020-08-28, 13:49:31 & 355.6977, 15.3505 & 13:32:17 & Setting & 88 \\
\hline
\multirow{6}{*}{
\begin{tabular}{l}
NuSTAR/FPM\\(10602002008)
\end{tabular}}
& 2020-08-29, 04:54:34 & 301.0417, 3.4484 & 00:58:42 & Rising & 89 \\
& 2020-08-29, 05:55:45 & 113.4503, 15.2920 & 13:29:32 & Setting & 90 \\
& 2020-08-29, 06:31:11 & 276.8268, 3.4309 & 00:58:28 & Rising & 89 \\
& 2020-08-29, 07:32:22 & 89.2221, 15.2732 & 13:29:15  & Setting & 90\\
& 2020-08-29, 08:07:49 & 252.6215, 3.3880 & 00:58:18 & Rising & 89 \\
& 2020-08-29, 09:09:00 & 64.9998, 15.2813 & 13:28:59  & Setting & 90 \\
\hline
\multirow{6}{*}{
\begin{tabular}{l}
NuSTAR/FPM\\(10702303002)
\end{tabular}}
& 2021-02-24, 02:02:25 & 162.7668, 15.4134 & 12:53:29 & Rising & 91 \\
& 2021-02-24, 03:03:38 & 335.4311, 3.9672 & 01:25:21 & Setting & 92\\
& 2021-02-24, 03:39:02 & 138.5451, 15.4127 & 12:53:12 & Rising & 91 \\
& 2021-02-24, 04:40:16 & 311.2310, 4.0295 & 01:25:10  & Setting & 92\\
& 2021-02-24, 05:15:39 & 114.3234, 15.4115 & 12:52:55 & Rising & 91 \\
& 2021-02-24, 06:16:53 & 287.0255, 4.0638 & 01:24:59 & Setting & 92\\
\hline
\multirow{6}{*}{
\begin{tabular}{l}
NuSTAR/FPM\\(10702303004)
\end{tabular}}
& 2021-02-24, 19:45:11 & 256.3431, 15.4164 & 12:50:32 & Rising & 93 \\
& 2021-02-24, 20:46:28 & 69.1889, 4.4436 & 01:23:12 & Setting & 94 \\
& 2021-02-24, 21:21:48 & 232.1256, 15.4122 & 12:50:17 & Rising & 93 \\
& 2021-02-24, 22:23:05 & 44.9830, 4.4811 & 01:22:59 & Setting & 94 \\
& 2021-02-24, 22:58:24 & 207.9022, 15.4366 & 12:49:59 & Rising & 93 \\
& 2021-02-24, 23:59:42 & 20.7811, 4.5173 & 01:22:49  & Setting & 94 \\
\hline
\multirow{6}{*}{
\begin{tabular}{l}
NuSTAR/FPM\\(10702303006)
\end{tabular}}
& 2021-08-28, 15:50:19 & 135.8574, 8.4183 & 00:53:44 & Rising & 95 \\
& 2021-08-28, 16:50:49 & 305.8260, 4.1961 & 13:14:07 & Setting & 96 \\
& 2021-08-28, 17:26:55 & 111.6184, 8.4748 & 00:53:22 & Rising & 95 \\
& 2021-08-28, 18:27:25 & 281.5903, 4.1548 & 13:13:45 & Setting & 96\\
& 2021-08-28, 19:03:31 & 87.3795, 8.5313 &  00:53:02 & Rising & 95 \\
& 2021-08-28, 20:04:01 & 257.3587, 4.1121 & 13:13:27 & Setting & 96\\
\hline
\multirow{6}{*}{
\begin{tabular}{l}
NuSTAR/FPM\\(10702303008)
\end{tabular}}
& 2021-08-29, 07:56:20 & 253.4673, 8.9665 & 00:50:12 & Rising & 97 \\
& 2021-08-29, 08:56:51 & 63.5047, 3.8224 &  13:10:51 & Setting & 98 \\
& 2021-08-29, 09:32:56 & 229.2247, 9.0214 & 00:49:49 & Rising & 97 \\
& 2021-08-29, 10:33:27 & 39.2736, 3.7825 &  13:10:32 & Setting & 98 \\
& 2021-08-29, 11:09:32 & 204.9864, 9.0780 & 00:49:27 & Rising & 97 \\
& 2021-08-29, 12:10:03 & 15.0386, 3.7447 &  13:10:11 & Setting & 98 \\
\hline
\multirow{6}{*}{
\begin{tabular}{l}
NuSTAR/FPM\\(10802303002)
\end{tabular}}
& 2022-02-24, 02:33:19 & 160.0056, 4.0911 & 13:13:19 & Rising & 99 \\
& 2022-02-24, 03:34:33 & 332.6630, 15.2855 & 01:45:11 & Setting & 100 \\
& 2022-02-24, 04:09:54 & 135.8067, 4.0655 & 13:13:06 & Rising & 99 \\
& 2022-02-24, 05:11:08 & 308.4491, 15.2763 & 01:44:54 & Setting & 100 \\
& 2022-02-24, 05:46:30 & 111.6091, 4.0102 & 13:12:56 & Rising & 99 \\
& 2022-02-24, 06:47:43 & 284.2311, 15.2684 & 01:44:38 & Setting & 100 \\
\hline
\multirow{2}{*}{
\begin{tabular}{l}
Hitomi/HXI\\(100044010) 
\end{tabular}}
 & 2016-03-25, 14:52:43 & 144.4674, 36.2736 & 00:30:34 & Setting & 101 \\
 & 2016-03-25, 16:28:44 & 120.3037, 36.1348 & 00:29:56 & Setting & 101 \\
\end{longtable}

\clearpage

\end{document}